\documentclass[%
 reprint,
 superscriptaddress,
 amsmath,amssymb,
 aps,
 prl,
floatfix,
]{revtex4-2}

\usepackage[colorlinks=true,linkcolor=black,anchorcolor=black,citecolor=blue]{hyperref}
\newcommand{\ket}[1]{\left\vert #1 \right\rangle}

\usepackage[latin1]{inputenc}
\usepackage{bm}
\usepackage{multirow,amssymb,amsbsy,amsmath}
\usepackage{graphicx}
\usepackage{verbatim}
\usepackage{upgreek}
\usepackage[T1]{fontenc}
\makeatletter
\usepackage{pifont}
\makeatother

\begin{document}

\title{On-demand quantum storage of photonic qubits in an on-chip waveguide}

\author{Chao Liu}
\author{Tian-Xiang Zhu}
\author{Ming-Xu Su}
%\author{Tao Tu}
%\email{email:tutao@ustc.edu.cn}
%\author{Pei-Yun Li}
%\author{Xiao Liu}
%\author{Xing-Yu Zhu}
\author{You-Zhi Ma}
\author{Zong-Quan Zhou}
\email{email:zq\_zhou@ustc.edu.cn}
% \author{Zong-Feng Li}
% \author{Peng-Jun Liang}
\author{Chuan-Feng Li}
\email{email:cfli@ustc.edu.cn}
\author{Guang-Can Guo}
% \footnote{*These authors contribute equally to this work.}

\affiliation{CAS Key Laboratory of Quantum Information, University of Science and Technology of China, Hefei 230026, China\\}
\affiliation{CAS Center For Excellence in Quantum Information and Quantum Physics, University of Science and Technology of China, Hefei 230026, China \\}
\date{\today }

\begin{abstract}
{Photonic quantum memory is the core element in quantum information processing (QIP). For the scalable and convenient practical applications, great efforts have been devoted to the integrated quantum memory based on various waveguides fabricated in solids. However, on-demand storage of qubits, which is an essential requirement for QIP, is still challenging to be implemented using such integrated quantum memory. Here we report the on-demand storage of time-bin qubits in an on-chip waveguide memory fabricated on the surface of a $^{151}$Eu$^{3+}$:Y$_2$SiO$_5$ crystal, utilizing the  Stark-modulated atomic frequency comb protocol. A qubit storage fidelity of $99.3\%\pm0.2\%$ is obtained with single-photon-level coherent pulses, far beyond the highest fidelity achievable using the classical measure-and-prepare strategy. The developed integrated quantum memory with the on-demand retrieval capability represents an important step toward practical applications of integrated quantum nodes in quantum networks.} 
\end{abstract}

\maketitle
Photonic quantum memory plays an important role in quantum information processing (QIP). Typical applications include enabling the long-distance quantum communication based on the quantum repeater approach \cite{gisin2011rmp,tittel2009np,2001DLCZ}, enhancing the precision of quantum metrology \cite{gottesman2012longer, komar2014quantum}, converting the heralded photons to on-demand photons \cite{nunn2013enhancing} and synchronizing the operations in quantum computation \cite{qc2001nature,tittel2009np}. 
Quantum memory based on the integrated structure can further meet the requirements of ease of use, robustness and low consumption for scalable networking applications \cite{tit2011nature}. Photonic storage has been successfully implemented in the integrated quantum memory based on various fabrication methods applied in rare-earth-ion (REI) -doped solids, such as femtosecond-laser micromachining (FLM) \cite{reid2016prap, reid2018optica, reid2019prl, zzq2020optica, zzq2020prap}, focused-ion-beam milled nanoresonators \cite{fara2017sci, zhong2015nanophotonic, zhong2017interfacing, craiciu2019nanophotonic} and lithium niobate waveguides \cite{tit2011nature, sinclair2016proposal, sinclair2014spectral,tit2019prap}.
 
%Quantum memory for heralded single photons, coherent optical memories with high fidelity are demonstrated in the femtosecond-laser micromachining (FLM) waveguides, utilzing different memory schemes and waveguides of different types \cite{reid2018,reid2019prl,zzq2020optica,zzq2020arxiv}. Quantum memory is also demonstrated in the titanium indiffused $\mathrm {LiNbO_3}$ \cite{tit2011nature} and focused-ion-beam milled nano-resonators \cite{fara2017sci}. 
%The ability of storing and retrieving qubits with high fidelity and post-determined time is a fundamental requirement for a quantum memory.

As a synchronizing tool in QIP, on-demand readout of photonic qubits is a fundamental requirement for quantum memory \cite{gisin2011rmp,tittel2009np}. A typical solution, as widely employed in bulk material, is reversibly transferring the optical collective excitation into the spin-wave excitation using strong control pulses \cite{afzelius2009multimode, jobez2015coherent, reid2015prl, yang2018multiplexed}.
However, filtering of such strong control pulses in the integrated waveguide structure is extremely challenging due to the overlapped spatial mode between single photons and control pulses. As a result, to date, the demonstrated integrated quantum storage based on REI-doped solids is limited to a preprogrammed delay, mostly based on the atomic frequency comb (AFC) protocol \cite{reid2018optica, reid2019prl,fara2017sci,craiciu2019nanophotonic,tit2011nature, sinclair2016proposal,sinclair2014spectral}. The only exception is the recent demonstration in a nanophotonic resonator fabricated in a $\mathrm{Nd}$:$\mathrm{YVO_4}$ crystal, in which a dynamically controlled shift of AFC delay up to 10 ns is obtained \cite{fara2017sci}. Unfortunately, the controlled shift is less than the duration of the readout pulse, which severely limits its practical applications.
%As a specific example, the time-bin qubits are robust for long-distance quantum information processing as they are not sensitive to the fluctuation in the transmitted channels \cite{tb1999prl,tb2002pra}. On-demand quantum memory for time-bin qubits have been implemented in a $\mathrm{Pr^{3+}:Y_2SiO_5}$ crystal with a storage time with tens $\mu$s.  In a $\mathrm{Ti:Tm:LiNbO_3}$ optical waveguide, quantum memory for time-bin qubits is also demonstrated with a predetermined storage time of several nanoseconds  \cite{tit2011nature}. On-demand storage of time-bin qubits with a controlled read-out time of tens nanoseconds is also demonstrated in a nanobeam optical cavity\cite{fara2017sci}.

Here we report the high-fidelity and on-demand storage of time-bin qubits in an integrated quantum memory with a controlled storage time exceeding 2 $\mu$s, far beyond the pulse width ($\sim$100 ns). The recently proposed Stark-modulated AFC protocol \cite{stefan2020arx} is implemented in a FLM on-chip waveguide to achieve the goal of noise-free on-demand photonic qubit storage. This protocol is further modified here to introduce two electric pulses with opposite directions to eliminate the dephasing caused by the unwanted inhomogeneity in Stark shifts \cite{SM}.

%%%%%%%%%%%% FIGURE 1 %%%%%%%%%%%%%%%%%%%%%%%%%%%%%%%%%%
\begin{figure*}[tb]
\centering
\includegraphics[width=0.95\textwidth]{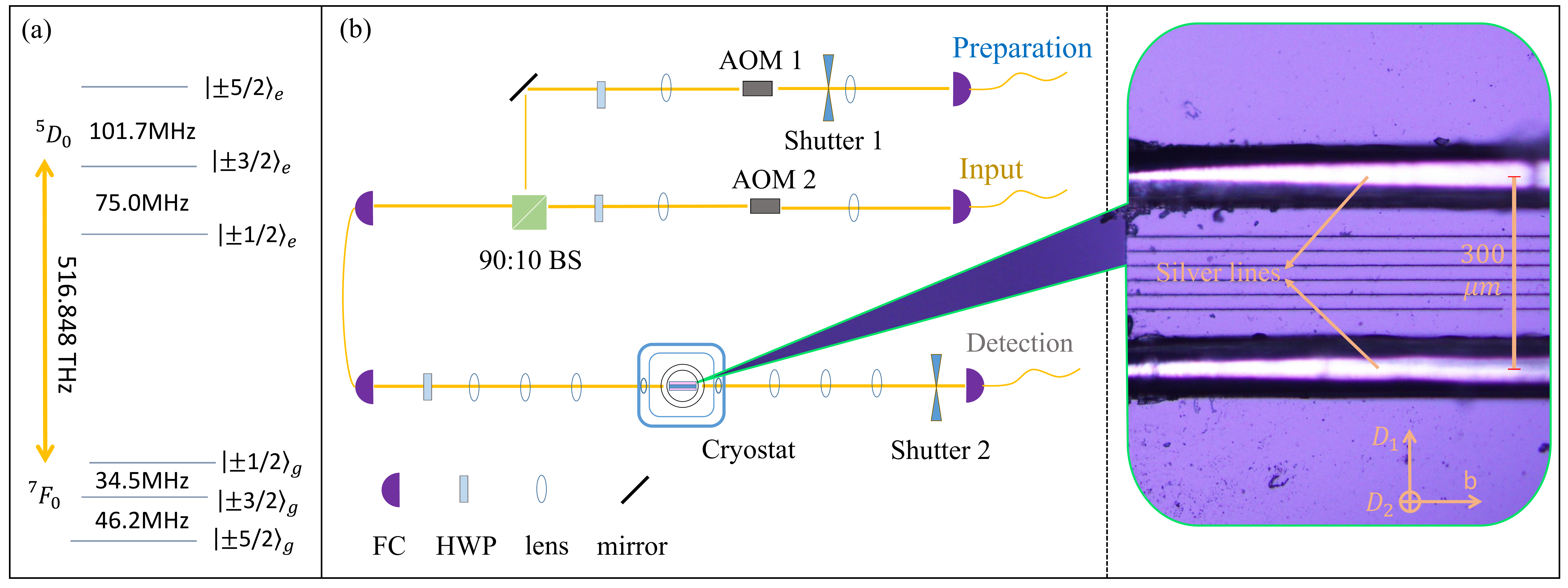}
\caption{\label{fig1} (color online) (a) Level structure of $^{151}$Eu$^{3+}$ ions at zero magnetic field. (b) Diagram of the experimental setup. The acoustic optic modulators labeled as AOM 1 and AOM 2 are employed to generate the preparation and input beams. The input and preparation beams are combined by a beam splitter (BS) with a reflection-to-transmission ratio of 90:10. The combined beam is coupled into the waveguide and then collected into a single-mode fiber with a lens group. The mechanical shutter 1 and shutter 2 ensure that the single-photon detector is protected from the strong preparation light. Inset: top view of the on-chip quantum memory under a microscope. Six tracks are fabricated on the sample with a spacing of 23 $\mu$m, forming five type \uppercase\expandafter{\romannumeral4} waveguides. The central one with the minimum insertion loss is employed for the quantum storage. Silver lines provide the electric field for storage time control. FC: Fiber coupler, HWP:half-wave plate.
}
\end{figure*}
%%%%%%%%%%%%%%%%%%%%%%%%%%%%%%%%%%%%%%%%%%%%%%%%%%%%%%%%

The AFC scheme has been one of the most successful protocols implemented in REI-doped solids with the advantages of wideband operation \cite{tit2011nature, saglamyurek2016multiplexed}, large multimode capacity \cite{reid2019prl, sinclair2014spectral, zhou2015quantum}, and high fidelity \cite{gisin2008nature, zhou2012realization}. The basic idea of this scheme is to prepare an absorption profile with a comb structure in the inhomogeneously broadened optical transition. An input photon will be absorbed by the AFC and the natural atomic dephasing will lead to a collective echo emission at the predetermined time of $1/\Delta$ due to the periodical absorption structure, where $\Delta$ is the periodicity of the comb \cite{gisin2008nature,afzelius2009multimode}. %Since $\Delta$ is pre-programmed in the preparation procedure, such two-level AFC scheme doesn't allow on-demand retrival of the photonic excitation. 
%When a photon is absorbed by the comb, it will be stored as a collective excitation of the atoms that resonate with the photon. The state can be written as 
%A photon absorbed by the comb results in a Dicke state \cite{gisin2008nature,gisin2009pra}:
%In bulk crystal, on-demand AFC memory can be demonstrated by applying two control pulses to map the coherence in and out the spin state \cite{gisin2009pra,reid2015prl}. Further more, the memory can works in the quantum regime by combining temporal, spectral and spatial filtering \cite{reid2015prl}, which is hard to be implemented in the integrated quantum memories \cite{reid2019prl,zzq2020optica}. 

Originated in \cite{lauritzen2011approaches}, the electric field gradient has been introduced to AFC protocol aiming at on-demand readout in discrete steps. Recently, Horvath et al. \cite{stefan2020arx} proposed a Stark-modulated AFC protocol, in which two homogeneous electric field pulses are employed %to actively control the rephasing of the atoms 
for storage time control, and a high signal-to-noise ratio (SNR) storage is demonstrated 
in a $\mathrm{Pr^{3+}}$:$\mathrm{Y_2SiO_5}$ crystal. Since no strong control optical pulses are required here, this scheme is suitable for implementations in the integrated quantum memory.
%Here, we replace the two bright control pulse with two electric pulse, inducing controlled reversible Stark shift to achieve on-demand AFC quantum memory.

For the $\mathrm{Y_2SiO_5}$ crystal, when the electric field is applied along the crystallographic $b$ axis or in the mirror plane, the Stark effect splits two group ions that acquire frequency shifts of the same magnitudes but opposite directions \cite{stark1992prb,stefan2020arx}. The magnitude of the frequency shift is $\Omega=(\bm{\mu_g}-\bm{\mu_e})\cdot \bm{E}/\hbar =\bm{\delta\mu} \cdot \bm{E}/\hbar$, where $\bm{\delta\mu}$ is the difference between the ground state electric-dipole moment $\bm{\mu_g}$ and the excited state electric-dipole moment $\bm{\mu_e}$, $\bm{E}$ is the applied electric field, $\hbar$ is the Planck constant. 
 %$For the $\mathrm{Y_2SiO_5}$ crystal, the Stark effect is not vanishing when the electric field is applied along the crystallographic b axis \cite{stefan2020arx} or in the mirror plane (perpendicular to crystallographic b axis), due to the space group symmetry of this crystal \cite{stark1992prb}. The electric filed 

 %Suppose the electric field is highly homogeneous along the transmission path of the input mode,  And 
Two groups of ions acquire the phases of $e^{i2\pi\bm{\Omega}T_p}$ and $e^{-i2\pi\bm{\Omega}T_p}$, respectively, for an electric field duration of $T_p$.
%  Then the Eq. (1) is modified to \cite{stark1992prb,stefan2020arx}:
%  \begin{equation}
%      \begin{split}
%               \left |\psi(t)\right \rangle=\sum\limits_{j=1}^{N^+}c_{j}^{+} e^{i2\pi\bm{\Omega}T}e^{-i2\pi\delta_{j}^{+} t} e^{-ikz_{j}^{+}}\left | g_1 \dots e_j \dots g_{N^{+}} \right \rangle \\ 
%  +\sum\limits_{j=1}^{N^-}c_{j}^{-}e^{-i2\pi\bm{\Omega}T} e^{-i2\pi\delta_{j}^{-} t} e^{-ikz_{j}^{-}}\left | g_1 \dots e_j \dots g_{N^{-}} \right \rangle. 
%      \end{split}
%  \end{equation}
% $N^{+}$ and $N^{-}$ is the total number of the ions that acquire the frequency shift of $\bm{\Omega}$ ($-\bm{\Omega}$). 
Considering the scenario when $T_p=1/(4{\Omega})$, the two groups of ions will accumulate a relative phase of $\pi$, inducing a destructive interference, thus the original AFC echo at $t=1/\Delta$ is effectively suppressed by the first electric pulse \cite{stefan2020arx}. A second electric pulse applied at time window $((n-1)/\Delta,n/\Delta)$ cancels the relative phase and leads to the on-demand retrieval of the echo at $n/\Delta$ where $n$ is an integer and $n\ge 2$. 
%If the electric field is completely homogeneous along the transmission path of the input mode, 
The storage efficiency is \cite{afzelius2009multimode,stefan2020arx}:
% \begin{equation}
% \begin{split}
%         \eta=1.13\frac{OD^2}{F^2}\exp({-1.06\frac{OD}{F}})\exp({-7.12\gamma^2t^2})
%         %\eta=\frac{\pi \cdot OD^2}{4\ln2 \cdot F^2}\exp(-\frac{OD}{2F}\sqrt{\frac{\pi}{\ln2}})\exp({-\frac{\pi^2\Delta^2}{2\ln2 \cdot F^2}t^2}) \\
%     %=1.13\frac{OD^2}{F^2}\exp({-1.06\frac{OD}{F}})\exp({-7.12\frac{\Delta^2}{F^2}t^2})
%     ,
% \end{split}
% \end{equation}
\begin{equation}
%\begin{split}
        \eta={\tilde{d}}^{\ 2}e^{-\tilde{d}}e^{-7.12\gamma^2t^2}
        %\eta=\frac{\pi \cdot OD^2}{4\ln2 \cdot F^2}\exp(-\frac{OD}{2F}\sqrt{\frac{\pi}{\ln2}})\exp({-\frac{\pi^2\Delta^2}{2\ln2 \cdot F^2}t^2}) \\
    %=1.13\frac{OD^2}{F^2}\exp({-1.06\frac{OD}{F}})\exp({-7.12\frac{\Delta^2}{F^2}t^2})
    ,
%\end{split}
\label{eq1}
\end{equation}
where $\tilde{d}$ is the effective absorption depth of the comb 
%where $OD$ is the optical depth of comb peaks, $F=\Delta/\gamma$ is the finesse of the AFC, 
and $\gamma$ is the full width at half maximum (FWHM) of the single absorption peak. 

In practice, the absorption peaks in the prepared AFC can be significantly broadened when the electric field is not homogeneous, this is the case in our experiments, as shown later. This electrically induced broadening causes an extra dephasing process during the total pulse duration ($2T_p$) of two pulses. A similar mechanism has been discussed in Ref. \cite{goldner2016pra}.
% %Considering the electric field as a function of the position $r$, the frequency shift of the ions can be written as ${\Omega}(r)=\bm{\delta\mu} \cdot \bm{E}(r)/\hbar$. %Each ion acquire a unique phase of $\Delta \phi=4{\Omega}(r)T_p$ after applying two positive electric pulses. %The storage efficiency can be greatly reduced due to the extra dephasing process induced by the inhomogeneity in Stark shifts. 
% As a toy model, here we assume a Gaussian distribution of the inhomogeneous broadening with a FWHM of $\gamma_I$, then
%   %\gamma_{\mathrm{E}}$, of the electric field can be described as Gaussian distribution, 
%     %of the frequency shift of the ions is  ${\gamma}=\bm{\delta\mu} \cdot \bm{\gamma_{\mathrm{E}}}/\hbar$. 
%   the storage efficiency is modified to \cite{beavan2012photon}
%  \begin{equation}
%  \eta'=\eta\exp(-2\frac{{\gamma_{I}^{2}{T_{p}^{2}}{\pi^{2}}}}{\ln(2)}) .
%       %{I}=I(0)\exp(-\frac{{\gamma}^{2}}{T_{p}^{2}}{\pi^{2}}} {2\ln(2)})
% \label{eq3}
% \end{equation}
% %where $I(0)$ is the stark modulated AFC echo intensity when the applied electric field is homogeneous. 
% The echo can be significantly suppressed when $\gamma_I$ approaches $1/T_p$. 
In our case, this broadening already leads to strongly suppressed signal as compared to that expected from Eq. (\ref{eq1}).
%  The decoherence time is characterized by the  reciprocal of the broadening caused by the field inhomogeneity. 
% In contrast to protocol in Ref. \cite{stephan}, 
Inspired by previous works \cite{hedges2010efficient,goldner2016pra}, here we apply the second electric field pulse with the same magnitude but the opposite direction with respect to the first pulse. 
%The dephasing caused by the inhomogeneity of Stark shifts can be completely rephased and the storage efficiency can be well described by the Eq. (1) \cite{SM}. 

The experimental sample is a $\mathrm{^{151}Eu^{3+}}$:$\mathrm{Y_2SiO_5}$ crystal with an isotope enrichment of $99\%$ for $^{151}$Eu. The substrate crystal has a dimension of $15 \times 5 \times 4$ mm$^3$ along the crystal's $b \times D_1 \times D_2$ axes and a dopant concentration of $0.1\%$. A series of type \uppercase\expandafter{\romannumeral4} waveguides are fabricated on the surface of the crystal utilizing the FLM system \cite{zzq2020prap}. %The type \uppercase\expandafter{\romannumeral4} waveguides are chosen here since they have the advantage of being conveniently integrated with other on-chip devices. %The type \uppercase\expandafter{\romannumeral4} waveguides support single-mode transmission of the input mode along the b axis, with polarization parallel to $D_1$ axis of the crystal. 
%Our five type \uppercase\expandafter{\romannumeral4} waveguides composed by six parallel ridges are fabricated by FLM system. 
Compared to our previous result presented in Ref. \cite{zzq2020prap}, now we find better fabricating parameters. The 1030 nm femtosecond laser has a pulse duration of 210 fs and a repetition rate of 201.9 kHz. A $50\times$ objective (NA=0.65) is utilized to focus the laser beam. Here we choose the low pulse energy of 60 nJ and adopt the multiscan method \cite{multiscan2007apl,multiscan20011oe} to minimize the propagation loss in the waveguide. The chosen waveguide has an end-to-end device efficiency of $40\%$, as defined by the ratio between the output of the single-mode fiber (SMF) and the input before the cryostat. This high device efficiency sets a new record for integrated photonic memories based on REI-doped solids. {The waveguide mode is close to the Gaussian mode and has a FWHM of $10.5\times7.9\ \mu$m \cite{SM}. As a comparison, the typical coupling efficiency is $10\%$ in the lithium niobate waveguide memory (waveguide mode size of $4.5\times3\ \mu$m FWHM) \cite{sinclair2014spectral,tit2011nature,tit2019prap,tittel2007prl}, while the coupling efficiency from the SMF to the focused-ion-beam milled waveguide memory is less than $27\%$ (fundamental mode volume of $0.0564\  \mu m^3$) \cite{zhong2016high, fara2017sci} }.% and the efficiency from free-space input to output is less than $24\%$ \cite{zhong2017interfacing, kindem2020control}.
% The width and the vertical depth of the type \uppercase\expandafter{\romannumeral4} waveguides are set as 21 $\mu$m and implement by multiple laser-writings at different depths beneath the surface of the sample because the vertical depth of the single track is only 7 $\mu$m. The total coupling efficiencies of the waveguides from the front of the cryostat to the single-mode fiber are $40\%$.

To introduce the required electric field for storage time control, two silver lines with a diameter of $80\ \mu$m and a spacing of $300\ \mu$m are buried in two grooves beside the waveguides (as shown in Fig. 1) which can apply electric field parallel to the $D_1$ axis. The grooves have a depth of approximately $80\ \mu$m and a width of about $100\ \mu$m, and are also fabricated with the FLM system. 

The experimental setup for quantum storage is similar to that presented in our previous work \cite{zzq2020optica}. The laser source is a frequency-doubled semiconductor laser which is externally locked at the frequency of 516.848 THz with a linewidth of subkilohertz. A cryostat (Montana Instrument) with a base temperature of 3.2 K is employed to cool down the sample. All optical pulses are modulated by the acoustic optic modulators (AOMs) which are driven by an arbitrary waveform generator. In addition, two mechanical shutters are placed in the pump path and before the single-photon detector (SPD), respectively, to protect the SPD in the preparation procedures. An arbitrary function generator (Tektronix, AFG3102C) %with a sampling rate of 1 GHz 
is utilized to directly generate the electric pulses.

%%%%%%%%%%%% FIGURE 2 %%%%%%%%%%%%%%%%%%%%%%%%%%%%%%%%%%
\begin{figure}[tb]
\centering
\includegraphics[width=0.5\textwidth]{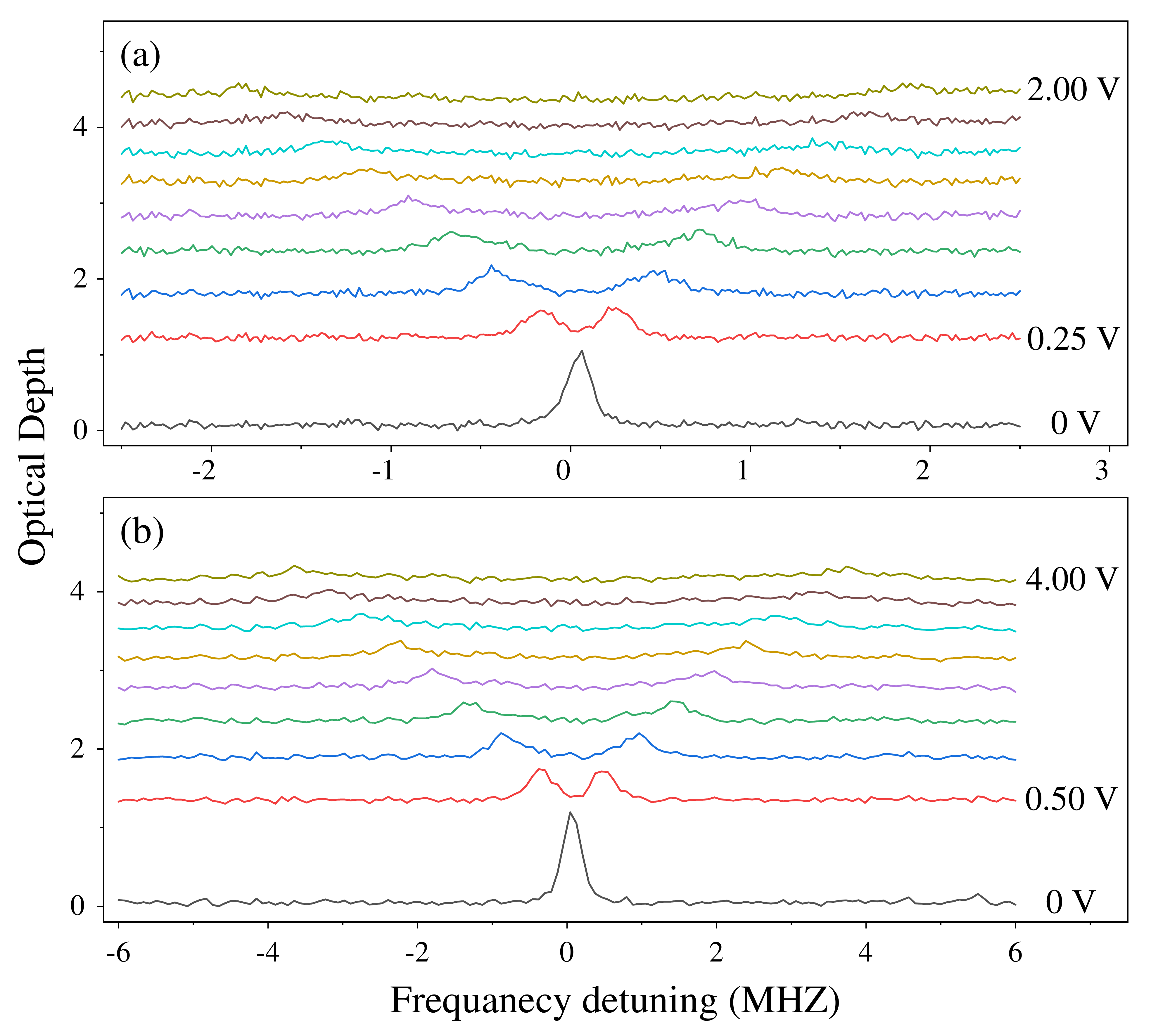}
\caption{\label{fig2} (color online) Spectral-hole burning measurements of the Stark coefficient of the $^7$F$_0$ to $^5$D$_0$ transition of $^{151}$Eu$^{3+}$ ions doped in Y$_2$SiO$_5$ crystal. The prepared absorption peak is split and broadened by the dc electric field. (a) A single class of ions is isolated and the probe laser is swept over a frequency range of 6 MHz. (b) No class-cleaning is implemented and the probe laser is swept over a frequency range of 12 MHz. %By utilizing a linear fitting of the shifts of antiholes, a Stark shift of approximately 0.92 MHz/V and 0.94 MHz/V are obtained for (a) and (b), respectively. %Only those traces with peaks that can be clearly identified are employed in the fitting procedures.
}
\end{figure}
%%%%%%%%%%%%%%%%%%%%%%%%%%%%%%%%%%%%%%%%%%%%%%%%%%%%%%%%

%%%%%%%%%%%% FIGURE 3 %%%%%%%%%%%%%%%%%%%%%%%%%%%%%%%%%%
\begin{figure}[tb]
\centering
\includegraphics[width=0.5\textwidth]{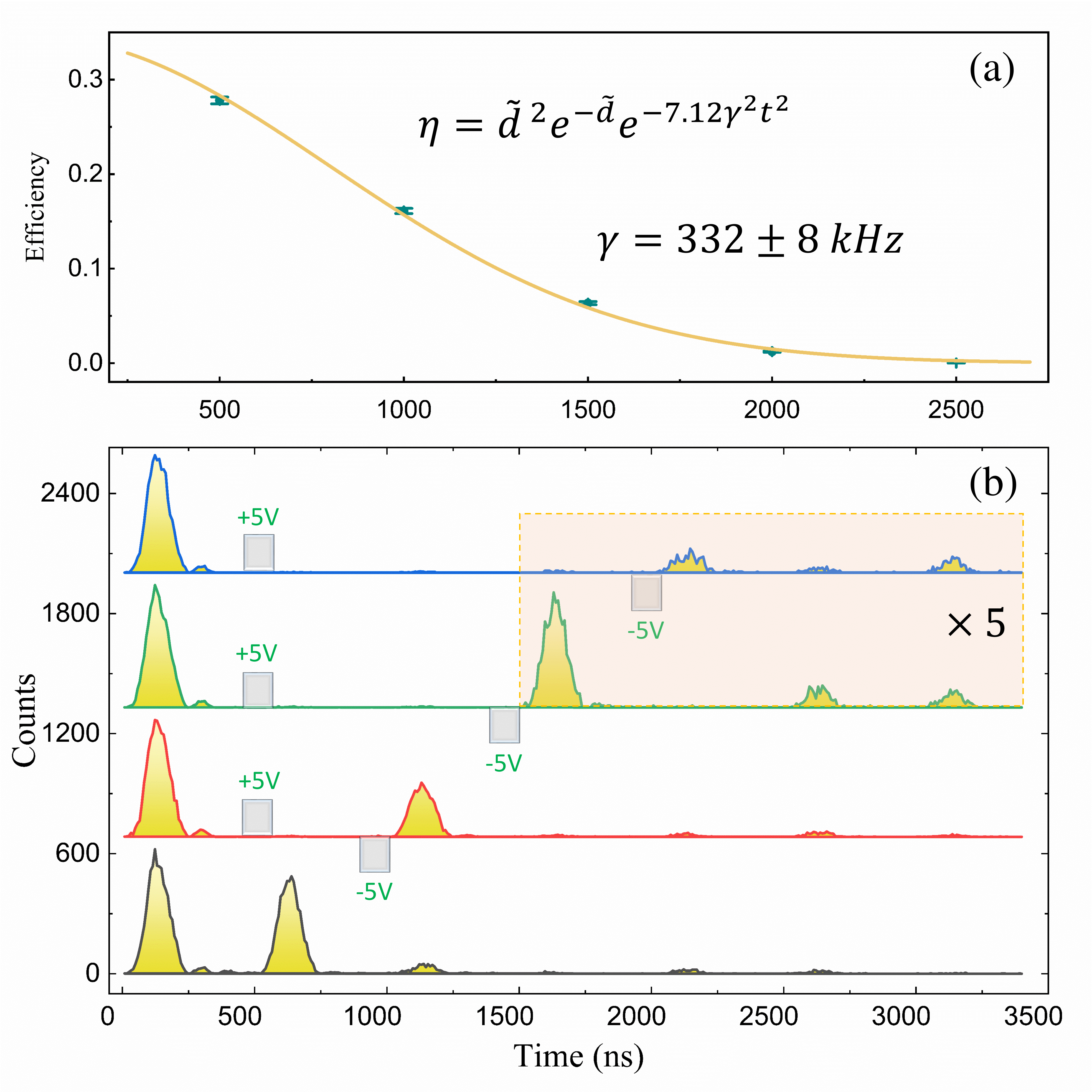}
\caption{\label{fig.3} (color online) (a) The storage efficiency as a function of the storage time. The peak width $\gamma$ is fitted as 332 $\pm$8 kHz using Eq. (1). (b) Photon counting histograms of on-demand storage with a weak coherent input of 0.5 photons per pulse. The AFC echo at the first order is completely suppressed after applying the first electric field pulse. After applying the second electric field pulse, the signal can be read out on-demand. The histograms in the pink shaded region are enlarged by 5 times. %The tails after the retrieved signal are caused by the bandwidth mismatching between the input mode and the AFC. 
}
\end{figure}
%%%%%%%%%%%%%%%%%%%%%%%%%%%%%%%%%%%%%%%%%%%%%%%%%%%%%%%%

%To our knowledge, the Stark coefficient for the $^7$F$_0$ to $^5$D$_0$ transition of the $\mathrm {Eu^{3+}:Y_2SiO_5}$ crystal has not been measured before. %A relevant measurement is implemented in an $\mathrm{Eu^{3+}:YAlO_3}$ crystal, giving an optical Stark coefficient of 33.7 kHz/Vcm$^{-1}$ \cite{stark1992prb} for this transition. 
We characterize the Stark effect of our sample using spectral-hole burning measurements. %The Eu$^{3+}$ ions are firstly initialized by sweeping the laser frequency around 200 MHz. %with a duration of 1 ms. This step is repeated for 300 times 
%so that any structure created in previous experiment cycle is completely erased. Then a transparent window is burned at the center frequency of $f_0$ with a bandwidth of 20 MHz %At last, two pulses at the center frequency of $f_0+34.5$ MHz and $f_0+5.7$ MHz 
% are injected, and with duration of 200 ms 
We create a single absorption peak in the center of a transparent spectral window.
%The minimum FWHM of the absorption peak here is limited by the optical coherence time $T_{2opt}$ of Eu$^{3+}$ ions, which is measured to approximately 50 $\mu$s here. This measured $T_{2opt}$ is significantly shorter than that we measured before \cite{zzq2020optica}. The possible reason may be that the actual temperature of the sample is higher because of the additional heating load from this device. 
The Stark coefficient can be obtained by looking on the frequency shift of the absorption peak in a dc bias electric field. Figure. 2(b) shows the linear shift of the peak when increasing the magnitude of the electric field. A linear fitting of the peak position of each peak gives a Stark coefficient of $28.2$ kHz/Vcm$^{-1}$.
%, which agrees with the value obtained by the electric modulated AFC echo measurement ($27.3$ kHz/Vcm$^{-1}$). 
Here, we do not pick up a single class of ions, i.e. various transitions between the hyperfine levels in the excited state and the ground state are involved \cite{lauritzen2012spectroscopic}. 
Then we conduct the spectral-hole burning measurements again, while first isolating a single class of ions using the pumping scheme described in Ref. \cite{zzq2020optica}. In this way, we specifically measure the Stark coefficient for the transition ${\left | \pm {1/2} \right \rangle}_g$ $\rightarrow$ ${\left | \pm {5/2} \right \rangle}_e$. The measured Stark coefficient is 28.6 kHz/Vcm$^{-1}$ (Fig. 2(a)) which is basically the same as that obtained in Fig. 2(b), considering the errors in identifying the peak centers of the broadened peaks. Our measurement agrees with the value reported in the previous work \cite{goldner2014prl}.

In order to increase the available optical depth for efficient storage, we pump the ions with transition frequency of ($f_0$-72, $f_0$-7) and ($f_0$+7, $f_0$+72) MHz, %with a sweeping duration of 1 ms, 
forming a strong absorption peak at $f_0$ with a bandwidth of 14 MHz and an optical depth of 5.9. %This is enhanced by 2 times as compared to the absorption depth after isolating a single class of ions. 
An AFC with a total bandwidth of 11 MHz and a comb periodicity of $\Delta=2$ MHz is prepared. On-demand retrieval of the AFC echo is enabled by applying two electric pulses with opposite directions as explained above. The electric pulse has a FWHM of 55 ns to induce a relative phase shift of $\pi$ between the two groups of ions to suppress the ordinary AFC echo. %The spacing between the two electric pulses actively control the storage time of the memory. 
{Fig. 3(b) presents the photon counting histograms of the on-demand AFC storage with a variable storage time up to 2 $\mu$s, with an input of weak coherent pulse with the average photon number $\mu_{in}=0.5$.} The input signal pulse has a FWHM of 100 ns. The efficiency for a storage time of 1 $\mu$s is $16.1\%\pm0.3\%$, while a SNR of $870\pm430$ is obtained \cite{SM}. %This agrees well with the the theoretical value of $15.7\%\pm0.1\%$. 
{The storage efficiency as a function of the storage time (as shown in Fig. 3(a)) is well fitted by Eq. (1) with $\gamma=332\pm8$ kHz and $\tilde{d}=0.94\pm0.03$, demonstrating the dephasing caused by the inhomogeneity of Stark shifts is completely eliminated.}
%**Thereotical estimations included here**
%%%%%%%%%%%% FIGURE 4 %%%%%%%%%%%%%%%%%%%%%%%%%%%%%%%%%%

\begin{figure}[tb]
\centering
\includegraphics[width=0.5\textwidth]{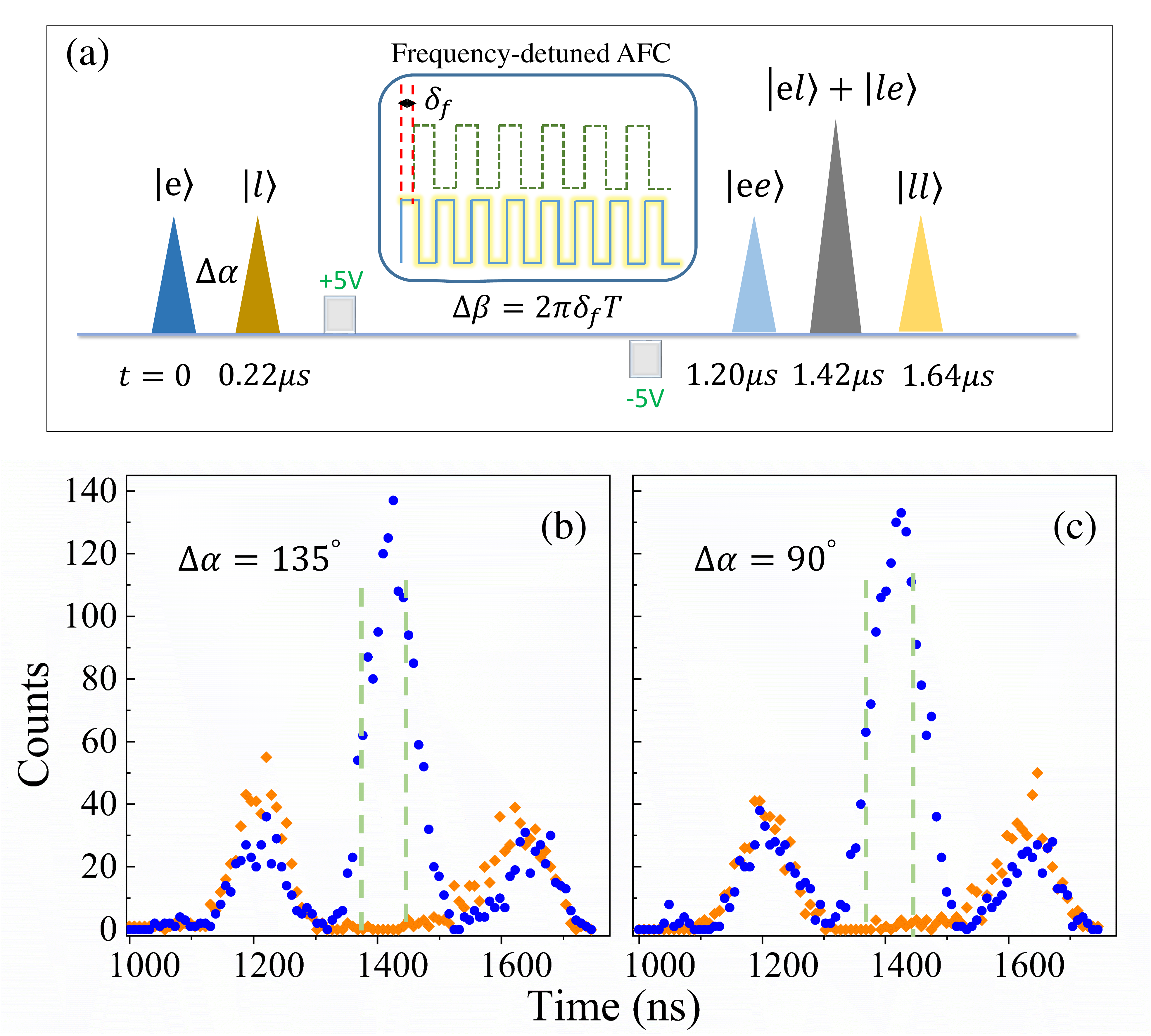}
\caption{\label{fig.4} (color online) (a) Sequence for storing, retrieving and analyzing the time-bin qubits. Inset: Illustration of the dynamical phase introduced by the frequency-detuned AFC. (b), (c) The constructive (blue dot) and destructive interference (orange square) fringes for the input state $\ket{e}+e^{i\Delta\alpha}\ket{l}$ with a relative phase of (b) $\Delta\alpha=135^{\circ}$ and (c) $\Delta\alpha=90^\circ$. The visibility of $99.6\%\pm0.2\%$ and $99.1\%\pm0.3\%$ are obtained respectively, with an integration window of 100 ns which is marked by a pair of green dashed lines.}
\end{figure}
%%%%%%%%%%%%%%%%%%%%%%%%%%%%%%%%%%%%%%%%%%%%%%%%%%%%%%%%

\begin{table*}[htbp]
  \centering
  \caption{Storage fidelities for various input states.}
  \scalebox{1.2}{
    \begin{tabular}{ccccccr}
    \toprule
    Input state & $\ket{e}$ & $\ket{l}$& $\ket{e}+i\ket{l}$   &$\ket{e}+e^{i3\pi/4}\ket{l}$ &$F_{T}$ \\
	\hline
     Fidelity &  $99.0\%\pm0.5\% $ &  $99.1\%\pm0.5\%$ &  $99.1\%\pm0.3\%$ & $99.6\%\pm0.2\%$&  $99.3\%\pm0.2\%$ \\
	\hline
    \end{tabular}
    }
  \label{tab:addlabel}
\end{table*}
%%%%%%%%%%%%%%%%%%%%%%%%%%%%%%%%%%%%%%%%
The above results demonstrate high-SNR storage of single-photon-level coherent states in the integrated memory and this device is ready for the qubit storage. 
Here we choose the time-bin encoded qubits which are particularly robust for long-distance transmission in optical fibers \cite{tb1999prl,tb2002pra}. 
The analysis of time-bin qubits requires an unbalanced Mach-Zehnder interferometer (MZI) for projection on a superposition basis \cite{gisin2008nature,reid2015prl}. Here we rely on two superimposed AFC to serve as a built-in unbalanced MZI \cite{gisin2008nature} and a dynamical phase is introduced by the frequency-detuned AFC \cite{zzq2015lg} for projection on arbitrary superposition states \cite{SM}. 

The double AFC are programmed to give standard AFC storage times of 600 ns and 710 ns, respectively. Two single-photon-level coherent Gaussian pulses with a FWHM of 100 ns and a spacing of $\tau=220$ ns serve as input, generating a time-bin qubit with the form $\ket{\psi}=c_e\ket{e}+c_l e^{i\Delta\alpha}\ket{l}$ \cite{reid2015prl,gisin2008nature}, where $\ket{e}$ and $\ket{l}$ denote the early and late time bin, respectively. $\Delta\alpha$ is the relative phase between two pulses and $c_e^2+c_l^2=1$. As shown in Fig. 4(a), the memory outputs three time bins \{$\ket{ee},\ket{el}+\ket{le},\ket{ll}$\}, after applying the second electric pulse. 

%If the the memory process preserves the time-bin qubits, an interference will take place at the central time bin. 
The characterization of the $\ket{el}+\ket{le}$ %requires a variable phase control during the measurement process. This can be 
is accomplished by introducing a certain amount of frequency detuning $\delta_f$ of the AFC featuring delay of 1/$\Delta$= 600 ns. The phase introduced by the detuned AFC is $\Delta\beta=2\pi\delta_f T$ with a storage time of $T$ \cite{zzq2015lg}.
%$In each pump-probe cycle, 1500 pulses with $\mu_{in}=0.5$ and a pulse separation of 5 $\mu$s are stored in the memory. The complete pump-probe cycle is repeated for 200 times, leading to a total repetition of storage process of $3\times10^5$.

The fidelity for $\ket{e}$ and $\ket{l}$ is defined as $F_{e(l)}=\frac{S+N}{S+2N}$, while $N$ is the noise and $S$ is the signal without including the noise \cite{reid2015prl}. The fidelity of $\ket{e}$ and $\ket{l}$ is measured to be $F_{e}=99.0\%\pm0.5\%$ and $F_{l}=99.1\%\pm0.5\%$. We calculate the average fidelity as $F_{el}=99.1\%\pm{0.4\%}$. The fidelity of the coherence measurement is defined as $F=(1+V)/2$ \cite{gisin2003nature}. By varying the measurement phase $\Delta\beta$ through the frequency-detuned AFC, we measure the interference visibility for input states of $\ket{e}+i\ket{l}$ and $\ket{e}+e^{i3\pi/4}\ket{l}$, as shown in Fig. 4. The interference visibility is $V=\frac{max-min}{max+min}$, with $max$ and $min$ corresponding to the maximum and minimum photon counts measured with $\Delta\beta=-\Delta\alpha$ and $\Delta\beta=-\Delta\alpha+\pi$, respectively. The storage fidelity for $\ket{e}+i\ket{l}$ and $\ket{e}+e^{i3\pi/4}\ket{l}$ is $99.1\%\pm0.3\%$ and $99.6\%\pm0.2\%$, respectively, and the mean fidelity  $F_{+-}=99.4\%\pm0.2\%$.  

The total storage fidelity, as defined by $F_T=\frac{1}{3}F_{el}+\frac{2}{3}F_{+-}$ \cite{reid2015prl} is calculated to be $99.3\%\pm0.2\%$. As a comparison, one can calculate the best achievable fidelity using the classical measure-and-prepare strategy, taking into account the Poissonian statistics of the input and the finite storage efficiency \cite{bound2011nature,yang2018multiplexed}. Considering the average photon number of 0.5 and the storage efficiency of $1.1\%\pm0.1\%$ in double AFC measurements, the strict classical fidelity bound is $81.4\%$. The experimentally achieved fidelity violates the classical bound by more than 89 standard deviations, demonstrating its high reliability in qubit storage.

In conclusion, based on an on-chip waveguide memory combined with on-chip electrodes, we demonstrate on-demand storage of photonic time-bin qubits with the fidelity of $99.3\%\pm0.2\%$. This is approximately the same as the highest fidelity obtained in a quantum memory based on the bulk material \cite{zhou2012realization}. The small volume of the integrated memory enables the use of a transistor-transistor-logic-compatible voltage for actively controlling the storage time. In the current demonstration, the storage time is limited by the broadened absorption peaks. The storage time can be greatly extended with much narrower AFC peaks and is fundamentally limited by the optical coherence time \cite{SM}. This compact, reliable and low-insertion-loss integrated quantum memory shall enable plentiful applications in QIP, especially in the construction of large-scale quantum networks based on integrated quantum nodes.

{\bf  Acknowledgments}
Chao Liu and Tian-Xiang Zhu contributed equally to this work.
This work is supported by the National Key R\&D Program of China (Grant No. 2017YFA0304100), the National Natural Science Foundation of China (Grant No. 11774331, No. 11774335, No. 11504362, No. 11821404, and No. 11654002), the Anhui Initiative in Quantum Information Technologies (Grant No. AHY020100), the Key Research Program of Frontier Sciences, the CAS (Grant No. QYZDY-SSW-SLH003), the Science Foundation of the CAS (Grant No. ZDRW-XH-2019-1) and the Fundamental Research Funds for the Central Universities (Grant No. WK2470000026, and No. WK2470000029).
Z.-Q.Z acknowledges the support from the Youth Innovation Promotion Association CAS.

\bibliographystyle{apsrev4-2.bst}
\bibliography{ec-afc}

\providecommand{\noopsort}[1]{}\providecommand{\singleletter}[1]{#1}%
\begin{thebibliography}{47}%
\makeatletter
\providecommand \@ifxundefined [1]{%
 \@ifx{#1\undefined}
}%
\providecommand \@ifnum [1]{%
 \ifnum #1\expandafter \@firstoftwo
 \else \expandafter \@secondoftwo
 \fi
}%
\providecommand \@ifx [1]{%
 \ifx #1\expandafter \@firstoftwo
 \else \expandafter \@secondoftwo
 \fi
}%
\providecommand \natexlab [1]{#1}%
\providecommand \enquote  [1]{``#1''}%
\providecommand \bibnamefont  [1]{#1}%
\providecommand \bibfnamefont [1]{#1}%
\providecommand \citenamefont [1]{#1}%
\providecommand \href@noop [0]{\@secondoftwo}%
\providecommand \href [0]{\begingroup \@sanitize@url \@href}%
\providecommand \@href[1]{\@@startlink{#1}\@@href}%
\providecommand \@@href[1]{\endgroup#1\@@endlink}%
\providecommand \@sanitize@url [0]{\catcode `\\12\catcode `\$12\catcode
  `\&12\catcode `\#12\catcode `\^12\catcode `\_12\catcode `\%12\relax}%
\providecommand \@@startlink[1]{}%
\providecommand \@@endlink[0]{}%
\providecommand \url  [0]{\begingroup\@sanitize@url \@url }%
\providecommand \@url [1]{\endgroup\@href {#1}{\urlprefix }}%
\providecommand \urlprefix  [0]{URL }%
\providecommand \Eprint [0]{\href }%
\providecommand \doibase [0]{https://doi.org/}%
\providecommand \selectlanguage [0]{\@gobble}%
\providecommand \bibinfo  [0]{\@secondoftwo}%
\providecommand \bibfield  [0]{\@secondoftwo}%
\providecommand \translation [1]{[#1]}%
\providecommand \BibitemOpen [0]{}%
\providecommand \bibitemStop [0]{}%
\providecommand \bibitemNoStop [0]{.\EOS\space}%
\providecommand \EOS [0]{\spacefactor3000\relax}%
\providecommand \BibitemShut  [1]{\csname bibitem#1\endcsname}%
\let\auto@bib@innerbib\@empty
%</preamble>
\bibitem [{\citenamefont {Sangouard}\ \emph {et~al.}(2011)\citenamefont
  {Sangouard}, \citenamefont {Simon}, \citenamefont {De~Riedmatten},\ and\
  \citenamefont {Gisin}}]{gisin2011rmp}%
  \BibitemOpen
  \bibfield  {author} {\bibinfo {author} {\bibfnamefont {N.}~\bibnamefont
  {Sangouard}}, \bibinfo {author} {\bibfnamefont {C.}~\bibnamefont {Simon}},
  \bibinfo {author} {\bibfnamefont {H.}~\bibnamefont {De~Riedmatten}},\ and\
  \bibinfo {author} {\bibfnamefont {N.}~\bibnamefont {Gisin}},\ }\href
  {https://doi.org/10.1103/RevModPhys.83.33} {\bibfield  {journal} {\bibinfo
  {journal} {Rev. Mod. Phys}\ }\textbf {\bibinfo {volume} {83}},\ \bibinfo
  {pages} {33} (\bibinfo {year} {2011})}\BibitemShut {NoStop}%
\bibitem [{\citenamefont {Lvovsky}\ \emph {et~al.}(2009)\citenamefont
  {Lvovsky}, \citenamefont {Sanders},\ and\ \citenamefont
  {Tittel}}]{tittel2009np}%
  \BibitemOpen
  \bibfield  {author} {\bibinfo {author} {\bibfnamefont {A.~I.}\ \bibnamefont
  {Lvovsky}}, \bibinfo {author} {\bibfnamefont {B.~C.}\ \bibnamefont
  {Sanders}},\ and\ \bibinfo {author} {\bibfnamefont {W.}~\bibnamefont
  {Tittel}},\ }\href {https://doi.org/10.1038/nphoton.2009.231} {\bibfield
  {journal} {\bibinfo  {journal} {Nat. photonics}\ }\textbf {\bibinfo {volume}
  {3}},\ \bibinfo {pages} {706} (\bibinfo {year} {2009})}\BibitemShut {NoStop}%
\bibitem [{\citenamefont {Duan}\ \emph {et~al.}(2001)\citenamefont {Duan},
  \citenamefont {Lukin}, \citenamefont {Cirac},\ and\ \citenamefont
  {Zoller}}]{2001DLCZ}%
  \BibitemOpen
  \bibfield  {author} {\bibinfo {author} {\bibfnamefont {L.-M.}\ \bibnamefont
  {Duan}}, \bibinfo {author} {\bibfnamefont {M.~D.}\ \bibnamefont {Lukin}},
  \bibinfo {author} {\bibfnamefont {J.~I.}\ \bibnamefont {Cirac}},\ and\
  \bibinfo {author} {\bibfnamefont {P.}~\bibnamefont {Zoller}},\ }\href
  {https://doi.org/10.1038/35106500} {\bibfield  {journal} {\bibinfo  {journal}
  {Nature}\ }\textbf {\bibinfo {volume} {414}},\ \bibinfo {pages} {413}
  (\bibinfo {year} {2001})}\BibitemShut {NoStop}%
\bibitem [{\citenamefont {Gottesman}\ \emph {et~al.}(2012)\citenamefont
  {Gottesman}, \citenamefont {Jennewein},\ and\ \citenamefont
  {Croke}}]{gottesman2012longer}%
  \BibitemOpen
  \bibfield  {author} {\bibinfo {author} {\bibfnamefont {D.}~\bibnamefont
  {Gottesman}}, \bibinfo {author} {\bibfnamefont {T.}~\bibnamefont
  {Jennewein}},\ and\ \bibinfo {author} {\bibfnamefont {S.}~\bibnamefont
  {Croke}},\ }\href {https://doi.org/10.1103/PhysRevLett.109.070503} {\bibfield
   {journal} {\bibinfo  {journal} {Phys. Rev. Lett.}\ }\textbf {\bibinfo
  {volume} {109}},\ \bibinfo {pages} {070503} (\bibinfo {year}
  {2012})}\BibitemShut {NoStop}%
\bibitem [{\citenamefont {Komar}\ \emph {et~al.}(2014)\citenamefont {Komar},
  \citenamefont {Kessler}, \citenamefont {Bishof}, \citenamefont {Jiang},
  \citenamefont {S{\o}rensen}, \citenamefont {Ye},\ and\ \citenamefont
  {Lukin}}]{komar2014quantum}%
  \BibitemOpen
  \bibfield  {author} {\bibinfo {author} {\bibfnamefont {P.}~\bibnamefont
  {Komar}}, \bibinfo {author} {\bibfnamefont {E.~M.}\ \bibnamefont {Kessler}},
  \bibinfo {author} {\bibfnamefont {M.}~\bibnamefont {Bishof}}, \bibinfo
  {author} {\bibfnamefont {L.}~\bibnamefont {Jiang}}, \bibinfo {author}
  {\bibfnamefont {A.~S.}\ \bibnamefont {S{\o}rensen}}, \bibinfo {author}
  {\bibfnamefont {J.}~\bibnamefont {Ye}},\ and\ \bibinfo {author}
  {\bibfnamefont {M.~D.}\ \bibnamefont {Lukin}},\ }\href
  {https://doi.org/10.1038/nphys3000} {\bibfield  {journal} {\bibinfo
  {journal} {Nat. Phys.}\ }\textbf {\bibinfo {volume} {10}},\ \bibinfo {pages}
  {582} (\bibinfo {year} {2014})}\BibitemShut {NoStop}%
\bibitem [{\citenamefont {Nunn}\ \emph {et~al.}(2013)\citenamefont {Nunn},
  \citenamefont {Langford}, \citenamefont {Kolthammer}, \citenamefont
  {Champion}, \citenamefont {Sprague}, \citenamefont {Michelberger},
  \citenamefont {Jin}, \citenamefont {England},\ and\ \citenamefont
  {Walmsley}}]{nunn2013enhancing}%
  \BibitemOpen
  \bibfield  {author} {\bibinfo {author} {\bibfnamefont {J.}~\bibnamefont
  {Nunn}}, \bibinfo {author} {\bibfnamefont {N.}~\bibnamefont {Langford}},
  \bibinfo {author} {\bibfnamefont {W.}~\bibnamefont {Kolthammer}}, \bibinfo
  {author} {\bibfnamefont {T.}~\bibnamefont {Champion}}, \bibinfo {author}
  {\bibfnamefont {M.}~\bibnamefont {Sprague}}, \bibinfo {author} {\bibfnamefont
  {P.}~\bibnamefont {Michelberger}}, \bibinfo {author} {\bibfnamefont {X.-M.}\
  \bibnamefont {Jin}}, \bibinfo {author} {\bibfnamefont {D.}~\bibnamefont
  {England}},\ and\ \bibinfo {author} {\bibfnamefont {I.}~\bibnamefont
  {Walmsley}},\ }\href {https://doi.org/10.1103/PhysRevLett.110.133601}
  {\bibfield  {journal} {\bibinfo  {journal} {Phys. Rev. Lett.}\ }\textbf
  {\bibinfo {volume} {110}},\ \bibinfo {pages} {133601} (\bibinfo {year}
  {2013})}\BibitemShut {NoStop}%
\bibitem [{\citenamefont {Knill}\ \emph {et~al.}(2001)\citenamefont {Knill},
  \citenamefont {Laflamme},\ and\ \citenamefont {Milburn}}]{qc2001nature}%
  \BibitemOpen
  \bibfield  {author} {\bibinfo {author} {\bibfnamefont {E.}~\bibnamefont
  {Knill}}, \bibinfo {author} {\bibfnamefont {R.}~\bibnamefont {Laflamme}},\
  and\ \bibinfo {author} {\bibfnamefont {G.~J.}\ \bibnamefont {Milburn}},\
  }\href {https://doi.org/Doi 10.1038/35051009} {\bibfield  {journal} {\bibinfo
   {journal} {Nature}\ }\textbf {\bibinfo {volume} {409}},\ \bibinfo {pages}
  {46} (\bibinfo {year} {2001})}\BibitemShut {NoStop}%
\bibitem [{\citenamefont {Saglamyurek}\ \emph {et~al.}(2011)\citenamefont
  {Saglamyurek}, \citenamefont {Sinclair}, \citenamefont {Jin}, \citenamefont
  {Slater}, \citenamefont {Oblak}, \citenamefont {Bussieres}, \citenamefont
  {George}, \citenamefont {Ricken}, \citenamefont {Sohler},\ and\ \citenamefont
  {Tittel}}]{tit2011nature}%
  \BibitemOpen
  \bibfield  {author} {\bibinfo {author} {\bibfnamefont {E.}~\bibnamefont
  {Saglamyurek}}, \bibinfo {author} {\bibfnamefont {N.}~\bibnamefont
  {Sinclair}}, \bibinfo {author} {\bibfnamefont {J.}~\bibnamefont {Jin}},
  \bibinfo {author} {\bibfnamefont {J.~A.}\ \bibnamefont {Slater}}, \bibinfo
  {author} {\bibfnamefont {D.}~\bibnamefont {Oblak}}, \bibinfo {author}
  {\bibfnamefont {F.}~\bibnamefont {Bussieres}}, \bibinfo {author}
  {\bibfnamefont {M.}~\bibnamefont {George}}, \bibinfo {author} {\bibfnamefont
  {R.}~\bibnamefont {Ricken}}, \bibinfo {author} {\bibfnamefont
  {W.}~\bibnamefont {Sohler}},\ and\ \bibinfo {author} {\bibfnamefont
  {W.}~\bibnamefont {Tittel}},\ }\href {https://doi.org/10.1038/nature09719}
  {\bibfield  {journal} {\bibinfo  {journal} {Nature}\ }\textbf {\bibinfo
  {volume} {469}},\ \bibinfo {pages} {512} (\bibinfo {year}
  {2011})}\BibitemShut {NoStop}%
\bibitem [{\citenamefont {Corrielli}\ \emph {et~al.}(2016)\citenamefont
  {Corrielli}, \citenamefont {Seri}, \citenamefont {Mazzera}, \citenamefont
  {Osellame},\ and\ \citenamefont {De~Riedmatten}}]{reid2016prap}%
  \BibitemOpen
  \bibfield  {author} {\bibinfo {author} {\bibfnamefont {G.}~\bibnamefont
  {Corrielli}}, \bibinfo {author} {\bibfnamefont {A.}~\bibnamefont {Seri}},
  \bibinfo {author} {\bibfnamefont {M.}~\bibnamefont {Mazzera}}, \bibinfo
  {author} {\bibfnamefont {R.}~\bibnamefont {Osellame}},\ and\ \bibinfo
  {author} {\bibfnamefont {H.}~\bibnamefont {De~Riedmatten}},\ }\href
  {https://doi.org/10.1103/PhysRevApplied.5.054013} {\bibfield  {journal}
  {\bibinfo  {journal} {Phys. Rev. Appl.}\ }\textbf {\bibinfo {volume} {5}},\
  \bibinfo {pages} {054013} (\bibinfo {year} {2016})}\BibitemShut {NoStop}%
\bibitem [{\citenamefont {Seri}\ \emph {et~al.}(2018)\citenamefont {Seri},
  \citenamefont {Corrielli}, \citenamefont {Lago-Rivera}, \citenamefont
  {Lenhard}, \citenamefont {de~Riedmatten}, \citenamefont {Osellame},\ and\
  \citenamefont {Mazzera}}]{reid2018optica}%
  \BibitemOpen
  \bibfield  {author} {\bibinfo {author} {\bibfnamefont {A.}~\bibnamefont
  {Seri}}, \bibinfo {author} {\bibfnamefont {G.}~\bibnamefont {Corrielli}},
  \bibinfo {author} {\bibfnamefont {D.}~\bibnamefont {Lago-Rivera}}, \bibinfo
  {author} {\bibfnamefont {A.}~\bibnamefont {Lenhard}}, \bibinfo {author}
  {\bibfnamefont {H.}~\bibnamefont {de~Riedmatten}}, \bibinfo {author}
  {\bibfnamefont {R.}~\bibnamefont {Osellame}},\ and\ \bibinfo {author}
  {\bibfnamefont {M.}~\bibnamefont {Mazzera}},\ }\href
  {https://doi.org/10.1364/Optica.5.000934} {\bibfield  {journal} {\bibinfo
  {journal} {Optica}\ }\textbf {\bibinfo {volume} {5}},\ \bibinfo {pages} {934}
  (\bibinfo {year} {2018})}\BibitemShut {NoStop}%
\bibitem [{\citenamefont {Seri}\ \emph {et~al.}(2019)\citenamefont {Seri},
  \citenamefont {Lago-Rivera}, \citenamefont {Lenhard}, \citenamefont
  {Corrielli}, \citenamefont {Osellame}, \citenamefont {Mazzera},\ and\
  \citenamefont {de~Riedmatten}}]{reid2019prl}%
  \BibitemOpen
  \bibfield  {author} {\bibinfo {author} {\bibfnamefont {A.}~\bibnamefont
  {Seri}}, \bibinfo {author} {\bibfnamefont {D.}~\bibnamefont {Lago-Rivera}},
  \bibinfo {author} {\bibfnamefont {A.}~\bibnamefont {Lenhard}}, \bibinfo
  {author} {\bibfnamefont {G.}~\bibnamefont {Corrielli}}, \bibinfo {author}
  {\bibfnamefont {R.}~\bibnamefont {Osellame}}, \bibinfo {author}
  {\bibfnamefont {M.}~\bibnamefont {Mazzera}},\ and\ \bibinfo {author}
  {\bibfnamefont {H.}~\bibnamefont {de~Riedmatten}},\ }\href
  {https://doi.org/ARTN 08050210.1103/PhysRevLett.123.080502} {\bibfield
  {journal} {\bibinfo  {journal} {Phys. Rev. Lett.}\ }\textbf {\bibinfo
  {volume} {123}},\ \bibinfo {pages} {080502} (\bibinfo {year}
  {2019})}\BibitemShut {NoStop}%
\bibitem [{\citenamefont {Liu}\ \emph {et~al.}(2020)\citenamefont {Liu},
  \citenamefont {Zhou}, \citenamefont {Zhu}, \citenamefont {Zheng},
  \citenamefont {Jin}, \citenamefont {Liu}, \citenamefont {Li}, \citenamefont
  {Huang}, \citenamefont {Ma}, \citenamefont {Tu} \emph
  {et~al.}}]{zzq2020optica}%
  \BibitemOpen
  \bibfield  {author} {\bibinfo {author} {\bibfnamefont {C.}~\bibnamefont
  {Liu}}, \bibinfo {author} {\bibfnamefont {Z.-Q.}\ \bibnamefont {Zhou}},
  \bibinfo {author} {\bibfnamefont {T.-X.}\ \bibnamefont {Zhu}}, \bibinfo
  {author} {\bibfnamefont {L.}~\bibnamefont {Zheng}}, \bibinfo {author}
  {\bibfnamefont {M.}~\bibnamefont {Jin}}, \bibinfo {author} {\bibfnamefont
  {X.}~\bibnamefont {Liu}}, \bibinfo {author} {\bibfnamefont {P.-Y.}\
  \bibnamefont {Li}}, \bibinfo {author} {\bibfnamefont {J.-Y.}\ \bibnamefont
  {Huang}}, \bibinfo {author} {\bibfnamefont {Y.}~\bibnamefont {Ma}}, \bibinfo
  {author} {\bibfnamefont {T.}~\bibnamefont {Tu}}, \emph {et~al.},\ }\href
  {https://doi.org/10.1364/Optica.379166} {\bibfield  {journal} {\bibinfo
  {journal} {Optica}\ }\textbf {\bibinfo {volume} {7}},\ \bibinfo {pages} {192}
  (\bibinfo {year} {2020})}\BibitemShut {NoStop}%
\bibitem [{\citenamefont {Zhu}\ \emph {et~al.}(2020)\citenamefont {Zhu},
  \citenamefont {Liu}, \citenamefont {Zheng}, \citenamefont {Zhou},
  \citenamefont {Li},\ and\ \citenamefont {Guo}}]{zzq2020prap}%
  \BibitemOpen
  \bibfield  {author} {\bibinfo {author} {\bibfnamefont {T.-X.}\ \bibnamefont
  {Zhu}}, \bibinfo {author} {\bibfnamefont {C.}~\bibnamefont {Liu}}, \bibinfo
  {author} {\bibfnamefont {L.}~\bibnamefont {Zheng}}, \bibinfo {author}
  {\bibfnamefont {Z.-Q.}\ \bibnamefont {Zhou}}, \bibinfo {author}
  {\bibfnamefont {C.-F.}\ \bibnamefont {Li}},\ and\ \bibinfo {author}
  {\bibfnamefont {G.-C.}\ \bibnamefont {Guo}},\ }\href
  {https://doi.org/10.1103/PhysRevApplied.14.054071} {\bibfield  {journal}
  {\bibinfo  {journal} {Phys. Rev. Appl.}\ }\textbf {\bibinfo {volume} {14}},\
  \bibinfo {pages} {054071} (\bibinfo {year} {2020})}\BibitemShut {NoStop}%
\bibitem [{\citenamefont {Zhong}\ \emph
  {et~al.}(2017{\natexlab{a}})\citenamefont {Zhong}, \citenamefont {Kindem},
  \citenamefont {Bartholomew}, \citenamefont {Rochman}, \citenamefont
  {Craiciu}, \citenamefont {Miyazono}, \citenamefont {Bettinelli},
  \citenamefont {Cavalli}, \citenamefont {Verma}, \citenamefont {Nam} \emph
  {et~al.}}]{fara2017sci}%
  \BibitemOpen
  \bibfield  {author} {\bibinfo {author} {\bibfnamefont {T.}~\bibnamefont
  {Zhong}}, \bibinfo {author} {\bibfnamefont {J.~M.}\ \bibnamefont {Kindem}},
  \bibinfo {author} {\bibfnamefont {J.~G.}\ \bibnamefont {Bartholomew}},
  \bibinfo {author} {\bibfnamefont {J.}~\bibnamefont {Rochman}}, \bibinfo
  {author} {\bibfnamefont {I.}~\bibnamefont {Craiciu}}, \bibinfo {author}
  {\bibfnamefont {E.}~\bibnamefont {Miyazono}}, \bibinfo {author}
  {\bibfnamefont {M.}~\bibnamefont {Bettinelli}}, \bibinfo {author}
  {\bibfnamefont {E.}~\bibnamefont {Cavalli}}, \bibinfo {author} {\bibfnamefont
  {V.}~\bibnamefont {Verma}}, \bibinfo {author} {\bibfnamefont {S.~W.}\
  \bibnamefont {Nam}}, \emph {et~al.},\ }\href
  {https://doi.org/10.1126/science.aan5959} {\bibfield  {journal} {\bibinfo
  {journal} {Science}\ }\textbf {\bibinfo {volume} {357}},\ \bibinfo {pages}
  {1392} (\bibinfo {year} {2017}{\natexlab{a}})}\BibitemShut {NoStop}%
\bibitem [{\citenamefont {Zhong}\ \emph
  {et~al.}(2015{\natexlab{a}})\citenamefont {Zhong}, \citenamefont {Kindem},
  \citenamefont {Miyazono},\ and\ \citenamefont
  {Faraon}}]{zhong2015nanophotonic}%
  \BibitemOpen
  \bibfield  {author} {\bibinfo {author} {\bibfnamefont {T.}~\bibnamefont
  {Zhong}}, \bibinfo {author} {\bibfnamefont {J.~M.}\ \bibnamefont {Kindem}},
  \bibinfo {author} {\bibfnamefont {E.}~\bibnamefont {Miyazono}},\ and\
  \bibinfo {author} {\bibfnamefont {A.}~\bibnamefont {Faraon}},\ }\href
  {https://doi.org/10.1038/ncomms9206} {\bibfield  {journal} {\bibinfo
  {journal} {Nat. Commun.}\ }\textbf {\bibinfo {volume} {6}},\ \bibinfo {pages}
  {8206} (\bibinfo {year} {2015}{\natexlab{a}})}\BibitemShut {NoStop}%
\bibitem [{\citenamefont {Zhong}\ \emph
  {et~al.}(2017{\natexlab{b}})\citenamefont {Zhong}, \citenamefont {Kindem},
  \citenamefont {Rochman},\ and\ \citenamefont
  {Faraon}}]{zhong2017interfacing}%
  \BibitemOpen
  \bibfield  {author} {\bibinfo {author} {\bibfnamefont {T.}~\bibnamefont
  {Zhong}}, \bibinfo {author} {\bibfnamefont {J.~M.}\ \bibnamefont {Kindem}},
  \bibinfo {author} {\bibfnamefont {J.}~\bibnamefont {Rochman}},\ and\ \bibinfo
  {author} {\bibfnamefont {A.}~\bibnamefont {Faraon}},\ }\href
  {https://doi.org/10.1038/ncomms14107} {\bibfield  {journal} {\bibinfo
  {journal} {Nat. Commun.}\ }\textbf {\bibinfo {volume} {8}},\ \bibinfo {pages}
  {14107} (\bibinfo {year} {2017}{\natexlab{b}})}\BibitemShut {NoStop}%
\bibitem [{\citenamefont {Craiciu}\ \emph {et~al.}(2019)\citenamefont
  {Craiciu}, \citenamefont {Lei}, \citenamefont {Rochman}, \citenamefont
  {Kindem}, \citenamefont {Bartholomew}, \citenamefont {Miyazono},
  \citenamefont {Zhong}, \citenamefont {Sinclair},\ and\ \citenamefont
  {Faraon}}]{craiciu2019nanophotonic}%
  \BibitemOpen
  \bibfield  {author} {\bibinfo {author} {\bibfnamefont {I.}~\bibnamefont
  {Craiciu}}, \bibinfo {author} {\bibfnamefont {M.}~\bibnamefont {Lei}},
  \bibinfo {author} {\bibfnamefont {J.}~\bibnamefont {Rochman}}, \bibinfo
  {author} {\bibfnamefont {J.~M.}\ \bibnamefont {Kindem}}, \bibinfo {author}
  {\bibfnamefont {J.~G.}\ \bibnamefont {Bartholomew}}, \bibinfo {author}
  {\bibfnamefont {E.}~\bibnamefont {Miyazono}}, \bibinfo {author}
  {\bibfnamefont {T.}~\bibnamefont {Zhong}}, \bibinfo {author} {\bibfnamefont
  {N.}~\bibnamefont {Sinclair}},\ and\ \bibinfo {author} {\bibfnamefont
  {A.}~\bibnamefont {Faraon}},\ }\href
  {https://doi.org/10.1103/PhysRevApplied.12.024062} {\bibfield  {journal}
  {\bibinfo  {journal} {Phys. Rev. Appl.}\ }\textbf {\bibinfo {volume} {12}},\
  \bibinfo {pages} {024062} (\bibinfo {year} {2019})}\BibitemShut {NoStop}%
\bibitem [{\citenamefont {Sinclair}\ \emph {et~al.}(2016)\citenamefont
  {Sinclair}, \citenamefont {Heshami}, \citenamefont {Deshmukh}, \citenamefont
  {Oblak}, \citenamefont {Simon},\ and\ \citenamefont
  {Tittel}}]{sinclair2016proposal}%
  \BibitemOpen
  \bibfield  {author} {\bibinfo {author} {\bibfnamefont {N.}~\bibnamefont
  {Sinclair}}, \bibinfo {author} {\bibfnamefont {K.}~\bibnamefont {Heshami}},
  \bibinfo {author} {\bibfnamefont {C.}~\bibnamefont {Deshmukh}}, \bibinfo
  {author} {\bibfnamefont {D.}~\bibnamefont {Oblak}}, \bibinfo {author}
  {\bibfnamefont {C.}~\bibnamefont {Simon}},\ and\ \bibinfo {author}
  {\bibfnamefont {W.}~\bibnamefont {Tittel}},\ }\href
  {https://doi.org/10.1038/ncomms13454} {\bibfield  {journal} {\bibinfo
  {journal} {Nat. Commun.}\ }\textbf {\bibinfo {volume} {7}},\ \bibinfo {pages}
  {13454} (\bibinfo {year} {2016})}\BibitemShut {NoStop}%
\bibitem [{\citenamefont {Sinclair}\ \emph {et~al.}(2014)\citenamefont
  {Sinclair}, \citenamefont {Saglamyurek}, \citenamefont {Mallahzadeh},
  \citenamefont {Slater}, \citenamefont {George}, \citenamefont {Ricken},
  \citenamefont {Hedges}, \citenamefont {Oblak}, \citenamefont {Simon},
  \citenamefont {Sohler} \emph {et~al.}}]{sinclair2014spectral}%
  \BibitemOpen
  \bibfield  {author} {\bibinfo {author} {\bibfnamefont {N.}~\bibnamefont
  {Sinclair}}, \bibinfo {author} {\bibfnamefont {E.}~\bibnamefont
  {Saglamyurek}}, \bibinfo {author} {\bibfnamefont {H.}~\bibnamefont
  {Mallahzadeh}}, \bibinfo {author} {\bibfnamefont {J.~A.}\ \bibnamefont
  {Slater}}, \bibinfo {author} {\bibfnamefont {M.}~\bibnamefont {George}},
  \bibinfo {author} {\bibfnamefont {R.}~\bibnamefont {Ricken}}, \bibinfo
  {author} {\bibfnamefont {M.~P.}\ \bibnamefont {Hedges}}, \bibinfo {author}
  {\bibfnamefont {D.}~\bibnamefont {Oblak}}, \bibinfo {author} {\bibfnamefont
  {C.}~\bibnamefont {Simon}}, \bibinfo {author} {\bibfnamefont
  {W.}~\bibnamefont {Sohler}}, \emph {et~al.},\ }\href
  {https://doi.org/10.1103/PhysRevLett.113.053603} {\bibfield  {journal}
  {\bibinfo  {journal} {Phys. Rev. Lett.}\ }\textbf {\bibinfo {volume} {113}},\
  \bibinfo {pages} {053603} (\bibinfo {year} {2014})}\BibitemShut {NoStop}%
\bibitem [{\citenamefont {Askarani}\ \emph {et~al.}(2019)\citenamefont
  {Askarani}, \citenamefont {Puigibert}, \citenamefont {Lutz}, \citenamefont
  {Verma}, \citenamefont {Shaw}, \citenamefont {Nam}, \citenamefont {Sinclair},
  \citenamefont {Oblak},\ and\ \citenamefont {Tittel}}]{tit2019prap}%
  \BibitemOpen
  \bibfield  {author} {\bibinfo {author} {\bibfnamefont {M.~F.}\ \bibnamefont
  {Askarani}}, \bibinfo {author} {\bibfnamefont {M.~G.}\ \bibnamefont
  {Puigibert}}, \bibinfo {author} {\bibfnamefont {T.}~\bibnamefont {Lutz}},
  \bibinfo {author} {\bibfnamefont {V.~B.}\ \bibnamefont {Verma}}, \bibinfo
  {author} {\bibfnamefont {M.~D.}\ \bibnamefont {Shaw}}, \bibinfo {author}
  {\bibfnamefont {S.~W.}\ \bibnamefont {Nam}}, \bibinfo {author} {\bibfnamefont
  {N.}~\bibnamefont {Sinclair}}, \bibinfo {author} {\bibfnamefont
  {D.}~\bibnamefont {Oblak}},\ and\ \bibinfo {author} {\bibfnamefont
  {W.}~\bibnamefont {Tittel}},\ }\href {https://doi.org/ARTN 054056
  10.1103/PhysRevApplied.11.054056} {\bibfield  {journal} {\bibinfo  {journal}
  {Phys. Rev. Appl.}\ }\textbf {\bibinfo {volume} {11}},\ \bibinfo {pages}
  {054056} (\bibinfo {year} {2019})}\BibitemShut {NoStop}%
\bibitem [{\citenamefont {Afzelius}\ \emph {et~al.}(2009)\citenamefont
  {Afzelius}, \citenamefont {Simon}, \citenamefont {De~Riedmatten},\ and\
  \citenamefont {Gisin}}]{afzelius2009multimode}%
  \BibitemOpen
  \bibfield  {author} {\bibinfo {author} {\bibfnamefont {M.}~\bibnamefont
  {Afzelius}}, \bibinfo {author} {\bibfnamefont {C.}~\bibnamefont {Simon}},
  \bibinfo {author} {\bibfnamefont {H.}~\bibnamefont {De~Riedmatten}},\ and\
  \bibinfo {author} {\bibfnamefont {N.}~\bibnamefont {Gisin}},\ }\href
  {https://doi.org/10.1103/PhysRevA.79.052329} {\bibfield  {journal} {\bibinfo
  {journal} {Phys. Rev. A}\ }\textbf {\bibinfo {volume} {79}},\ \bibinfo
  {pages} {052329} (\bibinfo {year} {2009})}\BibitemShut {NoStop}%
\bibitem [{\citenamefont {Jobez}\ \emph {et~al.}(2015)\citenamefont {Jobez},
  \citenamefont {Laplane}, \citenamefont {Timoney}, \citenamefont {Gisin},
  \citenamefont {Ferrier}, \citenamefont {Goldner},\ and\ \citenamefont
  {Afzelius}}]{jobez2015coherent}%
  \BibitemOpen
  \bibfield  {author} {\bibinfo {author} {\bibfnamefont {P.}~\bibnamefont
  {Jobez}}, \bibinfo {author} {\bibfnamefont {C.}~\bibnamefont {Laplane}},
  \bibinfo {author} {\bibfnamefont {N.}~\bibnamefont {Timoney}}, \bibinfo
  {author} {\bibfnamefont {N.}~\bibnamefont {Gisin}}, \bibinfo {author}
  {\bibfnamefont {A.}~\bibnamefont {Ferrier}}, \bibinfo {author} {\bibfnamefont
  {P.}~\bibnamefont {Goldner}},\ and\ \bibinfo {author} {\bibfnamefont
  {M.}~\bibnamefont {Afzelius}},\ }\href
  {https://doi.org/10.1103/PhysRevLett.114.230502} {\bibfield  {journal}
  {\bibinfo  {journal} {Phys. Rev. Lett.}\ }\textbf {\bibinfo {volume} {114}},\
  \bibinfo {pages} {230502} (\bibinfo {year} {2015})}\BibitemShut {NoStop}%
\bibitem [{\citenamefont {G{\"u}ndo{\u{g}}an}\ \emph
  {et~al.}(2015)\citenamefont {G{\"u}ndo{\u{g}}an}, \citenamefont {Ledingham},
  \citenamefont {Kutluer}, \citenamefont {Mazzera},\ and\ \citenamefont
  {De~Riedmatten}}]{reid2015prl}%
  \BibitemOpen
  \bibfield  {author} {\bibinfo {author} {\bibfnamefont {M.}~\bibnamefont
  {G{\"u}ndo{\u{g}}an}}, \bibinfo {author} {\bibfnamefont {P.~M.}\ \bibnamefont
  {Ledingham}}, \bibinfo {author} {\bibfnamefont {K.}~\bibnamefont {Kutluer}},
  \bibinfo {author} {\bibfnamefont {M.}~\bibnamefont {Mazzera}},\ and\ \bibinfo
  {author} {\bibfnamefont {H.}~\bibnamefont {De~Riedmatten}},\ }\href
  {https://doi.org/10.1103/PhysRevLett.114.230501} {\bibfield  {journal}
  {\bibinfo  {journal} {Phys. Rev. Lett.}\ }\textbf {\bibinfo {volume} {114}},\
  \bibinfo {pages} {230501} (\bibinfo {year} {2015})}\BibitemShut {NoStop}%
\bibitem [{\citenamefont {Yang}\ \emph {et~al.}(2018)\citenamefont {Yang},
  \citenamefont {Zhou}, \citenamefont {Hua}, \citenamefont {Liu}, \citenamefont
  {Li}, \citenamefont {Li}, \citenamefont {Ma}, \citenamefont {Liu},
  \citenamefont {Liang}, \citenamefont {Li} \emph
  {et~al.}}]{yang2018multiplexed}%
  \BibitemOpen
  \bibfield  {author} {\bibinfo {author} {\bibfnamefont {T.-S.}\ \bibnamefont
  {Yang}}, \bibinfo {author} {\bibfnamefont {Z.-Q.}\ \bibnamefont {Zhou}},
  \bibinfo {author} {\bibfnamefont {Y.-L.}\ \bibnamefont {Hua}}, \bibinfo
  {author} {\bibfnamefont {X.}~\bibnamefont {Liu}}, \bibinfo {author}
  {\bibfnamefont {Z.-F.}\ \bibnamefont {Li}}, \bibinfo {author} {\bibfnamefont
  {P.-Y.}\ \bibnamefont {Li}}, \bibinfo {author} {\bibfnamefont
  {Y.}~\bibnamefont {Ma}}, \bibinfo {author} {\bibfnamefont {C.}~\bibnamefont
  {Liu}}, \bibinfo {author} {\bibfnamefont {P.-J.}\ \bibnamefont {Liang}},
  \bibinfo {author} {\bibfnamefont {X.}~\bibnamefont {Li}}, \emph {et~al.},\
  }\href {https://doi.org/10.1038/s41467-018-05669-5} {\bibfield  {journal}
  {\bibinfo  {journal} {Nat. Commun.}\ }\textbf {\bibinfo {volume} {9}},\
  \bibinfo {pages} {3407} (\bibinfo {year} {2018})}\BibitemShut {NoStop}%
\bibitem [{\citenamefont {Horvath}\ \emph {et~al.}(2020)\citenamefont
  {Horvath}, \citenamefont {Alqedra}, \citenamefont {Kinos}, \citenamefont
  {Walther}, \citenamefont {Kr{\"o}ll},\ and\ \citenamefont
  {Rippe}}]{stefan2020arx}%
  \BibitemOpen
  \bibfield  {author} {\bibinfo {author} {\bibfnamefont {S.~P.}\ \bibnamefont
  {Horvath}}, \bibinfo {author} {\bibfnamefont {M.~K.}\ \bibnamefont
  {Alqedra}}, \bibinfo {author} {\bibfnamefont {A.}~\bibnamefont {Kinos}},
  \bibinfo {author} {\bibfnamefont {A.}~\bibnamefont {Walther}}, \bibinfo
  {author} {\bibfnamefont {S.}~\bibnamefont {Kr{\"o}ll}},\ and\ \bibinfo
  {author} {\bibfnamefont {L.}~\bibnamefont {Rippe}},\ }\href@noop {}
  {\bibfield  {journal} {\bibinfo  {journal} {arXiv preprint arXiv:2006.00943}\
  } (\bibinfo {year} {2020})}\BibitemShut {NoStop}%
\bibitem [{SM()}]{SM}%
  \BibitemOpen
  \href@noop {} {\bibinfo {title} {See supplemental material at [url will be
  inserted by publisher] for details about the protocol and experiments, which
  includes refs. \cite{EuT2o,nature_6hour}.}}\BibitemShut {Stop}%
\bibitem [{\citenamefont {Saglamyurek}\ \emph {et~al.}(2016)\citenamefont
  {Saglamyurek}, \citenamefont {Puigibert}, \citenamefont {Zhou}, \citenamefont
  {Giner}, \citenamefont {Marsili}, \citenamefont {Verma}, \citenamefont {Nam},
  \citenamefont {Oesterling}, \citenamefont {Nippa}, \citenamefont {Oblak}
  \emph {et~al.}}]{saglamyurek2016multiplexed}%
  \BibitemOpen
  \bibfield  {author} {\bibinfo {author} {\bibfnamefont {E.}~\bibnamefont
  {Saglamyurek}}, \bibinfo {author} {\bibfnamefont {M.~G.}\ \bibnamefont
  {Puigibert}}, \bibinfo {author} {\bibfnamefont {Q.}~\bibnamefont {Zhou}},
  \bibinfo {author} {\bibfnamefont {L.}~\bibnamefont {Giner}}, \bibinfo
  {author} {\bibfnamefont {F.}~\bibnamefont {Marsili}}, \bibinfo {author}
  {\bibfnamefont {V.~B.}\ \bibnamefont {Verma}}, \bibinfo {author}
  {\bibfnamefont {S.~W.}\ \bibnamefont {Nam}}, \bibinfo {author} {\bibfnamefont
  {L.}~\bibnamefont {Oesterling}}, \bibinfo {author} {\bibfnamefont
  {D.}~\bibnamefont {Nippa}}, \bibinfo {author} {\bibfnamefont
  {D.}~\bibnamefont {Oblak}}, \emph {et~al.},\ }\href
  {https://doi.org/10.1038/ncomms11202} {\bibfield  {journal} {\bibinfo
  {journal} {Nat. Commun.}\ }\textbf {\bibinfo {volume} {7}},\ \bibinfo {pages}
  {11202} (\bibinfo {year} {2016})}\BibitemShut {NoStop}%
\bibitem [{\citenamefont {Zhou}\ \emph
  {et~al.}(2015{\natexlab{a}})\citenamefont {Zhou}, \citenamefont {Hua},
  \citenamefont {Liu}, \citenamefont {Chen}, \citenamefont {Xu}, \citenamefont
  {Han}, \citenamefont {Li},\ and\ \citenamefont {Guo}}]{zhou2015quantum}%
  \BibitemOpen
  \bibfield  {author} {\bibinfo {author} {\bibfnamefont {Z.-Q.}\ \bibnamefont
  {Zhou}}, \bibinfo {author} {\bibfnamefont {Y.-L.}\ \bibnamefont {Hua}},
  \bibinfo {author} {\bibfnamefont {X.}~\bibnamefont {Liu}}, \bibinfo {author}
  {\bibfnamefont {G.}~\bibnamefont {Chen}}, \bibinfo {author} {\bibfnamefont
  {J.-S.}\ \bibnamefont {Xu}}, \bibinfo {author} {\bibfnamefont {Y.-J.}\
  \bibnamefont {Han}}, \bibinfo {author} {\bibfnamefont {C.-F.}\ \bibnamefont
  {Li}},\ and\ \bibinfo {author} {\bibfnamefont {G.-C.}\ \bibnamefont {Guo}},\
  }\href {https://doi.org/10.1103/PhysRevLett.115.070502} {\bibfield  {journal}
  {\bibinfo  {journal} {Phys. Rev. Lett.}\ }\textbf {\bibinfo {volume} {115}},\
  \bibinfo {pages} {070502} (\bibinfo {year} {2015}{\natexlab{a}})}\BibitemShut
  {NoStop}%
\bibitem [{\citenamefont {De~Riedmatten}\ \emph {et~al.}(2008)\citenamefont
  {De~Riedmatten}, \citenamefont {Afzelius}, \citenamefont {Staudt},
  \citenamefont {Simon},\ and\ \citenamefont {Gisin}}]{gisin2008nature}%
  \BibitemOpen
  \bibfield  {author} {\bibinfo {author} {\bibfnamefont {H.}~\bibnamefont
  {De~Riedmatten}}, \bibinfo {author} {\bibfnamefont {M.}~\bibnamefont
  {Afzelius}}, \bibinfo {author} {\bibfnamefont {M.~U.}\ \bibnamefont
  {Staudt}}, \bibinfo {author} {\bibfnamefont {C.}~\bibnamefont {Simon}},\ and\
  \bibinfo {author} {\bibfnamefont {N.}~\bibnamefont {Gisin}},\ }\href {<Go to
  ISI>://WOS:000261559900045 https://www.nature.com/articles/nature07607.pdf}
  {\bibfield  {journal} {\bibinfo  {journal} {Nature}\ }\textbf {\bibinfo
  {volume} {456}},\ \bibinfo {pages} {773} (\bibinfo {year}
  {2008})}\BibitemShut {NoStop}%
\bibitem [{\citenamefont {Zhou}\ \emph {et~al.}(2012)\citenamefont {Zhou},
  \citenamefont {Lin}, \citenamefont {Yang}, \citenamefont {Li},\ and\
  \citenamefont {Guo}}]{zhou2012realization}%
  \BibitemOpen
  \bibfield  {author} {\bibinfo {author} {\bibfnamefont {Z.-Q.}\ \bibnamefont
  {Zhou}}, \bibinfo {author} {\bibfnamefont {W.-B.}\ \bibnamefont {Lin}},
  \bibinfo {author} {\bibfnamefont {M.}~\bibnamefont {Yang}}, \bibinfo {author}
  {\bibfnamefont {C.-F.}\ \bibnamefont {Li}},\ and\ \bibinfo {author}
  {\bibfnamefont {G.-C.}\ \bibnamefont {Guo}},\ }\href
  {https://doi.org/10.1103/PhysRevLett.108.190505} {\bibfield  {journal}
  {\bibinfo  {journal} {Phys. Rev. Lett.}\ }\textbf {\bibinfo {volume} {108}},\
  \bibinfo {pages} {190505} (\bibinfo {year} {2012})}\BibitemShut {NoStop}%
\bibitem [{\citenamefont {Lauritzen}\ \emph {et~al.}(2011)\citenamefont
  {Lauritzen}, \citenamefont {Min{\'a}{\v{r}}}, \citenamefont {De~Riedmatten},
  \citenamefont {Afzelius},\ and\ \citenamefont
  {Gisin}}]{lauritzen2011approaches}%
  \BibitemOpen
  \bibfield  {author} {\bibinfo {author} {\bibfnamefont {B.}~\bibnamefont
  {Lauritzen}}, \bibinfo {author} {\bibfnamefont {J.}~\bibnamefont
  {Min{\'a}{\v{r}}}}, \bibinfo {author} {\bibfnamefont {H.}~\bibnamefont
  {De~Riedmatten}}, \bibinfo {author} {\bibfnamefont {M.}~\bibnamefont
  {Afzelius}},\ and\ \bibinfo {author} {\bibfnamefont {N.}~\bibnamefont
  {Gisin}},\ }\href {https://doi.org/10.1103/PhysRevA.83.012318} {\bibfield
  {journal} {\bibinfo  {journal} {Phys. Rev. A}\ }\textbf {\bibinfo {volume}
  {83}},\ \bibinfo {pages} {012318} (\bibinfo {year} {2011})}\BibitemShut
  {NoStop}%
\bibitem [{\citenamefont {Meixner}\ \emph {et~al.}(1992)\citenamefont
  {Meixner}, \citenamefont {Jefferson},\ and\ \citenamefont
  {Macfarlane}}]{stark1992prb}%
  \BibitemOpen
  \bibfield  {author} {\bibinfo {author} {\bibfnamefont {A.}~\bibnamefont
  {Meixner}}, \bibinfo {author} {\bibfnamefont {C.}~\bibnamefont {Jefferson}},\
  and\ \bibinfo {author} {\bibfnamefont {R.}~\bibnamefont {Macfarlane}},\
  }\href {https://doi.org/10.1103/physrevb.46.5912} {\bibfield  {journal}
  {\bibinfo  {journal} {Phys. Rev. B}\ }\textbf {\bibinfo {volume} {46}},\
  \bibinfo {pages} {5912} (\bibinfo {year} {1992})}\BibitemShut {NoStop}%
\bibitem [{\citenamefont {Arcangeli}\ \emph {et~al.}(2016)\citenamefont
  {Arcangeli}, \citenamefont {Ferrier},\ and\ \citenamefont
  {Goldner}}]{goldner2016pra}%
  \BibitemOpen
  \bibfield  {author} {\bibinfo {author} {\bibfnamefont {A.}~\bibnamefont
  {Arcangeli}}, \bibinfo {author} {\bibfnamefont {A.}~\bibnamefont {Ferrier}},\
  and\ \bibinfo {author} {\bibfnamefont {P.}~\bibnamefont {Goldner}},\ }\href
  {https://doi.org/10.1103/PhysRevA.93.062303} {\bibfield  {journal} {\bibinfo
  {journal} {Phys. Rev. A}\ }\textbf {\bibinfo {volume} {93}},\ \bibinfo
  {pages} {062303} (\bibinfo {year} {2016})}\BibitemShut {NoStop}%
\bibitem [{\citenamefont {Hedges}\ \emph {et~al.}(2010)\citenamefont {Hedges},
  \citenamefont {Longdell}, \citenamefont {Li},\ and\ \citenamefont
  {Sellars}}]{hedges2010efficient}%
  \BibitemOpen
  \bibfield  {author} {\bibinfo {author} {\bibfnamefont {M.~P.}\ \bibnamefont
  {Hedges}}, \bibinfo {author} {\bibfnamefont {J.~J.}\ \bibnamefont
  {Longdell}}, \bibinfo {author} {\bibfnamefont {Y.}~\bibnamefont {Li}},\ and\
  \bibinfo {author} {\bibfnamefont {M.~J.}\ \bibnamefont {Sellars}},\ }\href
  {https://doi.org/10.1038/nature09081} {\bibfield  {journal} {\bibinfo
  {journal} {Nature}\ }\textbf {\bibinfo {volume} {465}},\ \bibinfo {pages}
  {1052} (\bibinfo {year} {2010})}\BibitemShut {NoStop}%
\bibitem [{\citenamefont {Osellame}\ \emph {et~al.}(2007)\citenamefont
  {Osellame}, \citenamefont {Lobino}, \citenamefont {Chiodo}, \citenamefont
  {Marangoni}, \citenamefont {Cerullo}, \citenamefont {Ramponi}, \citenamefont
  {Bookey}, \citenamefont {Thomson}, \citenamefont {Psaila},\ and\
  \citenamefont {Kar}}]{multiscan2007apl}%
  \BibitemOpen
  \bibfield  {author} {\bibinfo {author} {\bibfnamefont {R.}~\bibnamefont
  {Osellame}}, \bibinfo {author} {\bibfnamefont {M.}~\bibnamefont {Lobino}},
  \bibinfo {author} {\bibfnamefont {N.}~\bibnamefont {Chiodo}}, \bibinfo
  {author} {\bibfnamefont {M.}~\bibnamefont {Marangoni}}, \bibinfo {author}
  {\bibfnamefont {G.}~\bibnamefont {Cerullo}}, \bibinfo {author} {\bibfnamefont
  {R.}~\bibnamefont {Ramponi}}, \bibinfo {author} {\bibfnamefont
  {H.}~\bibnamefont {Bookey}}, \bibinfo {author} {\bibfnamefont
  {R.}~\bibnamefont {Thomson}}, \bibinfo {author} {\bibfnamefont
  {N.}~\bibnamefont {Psaila}},\ and\ \bibinfo {author} {\bibfnamefont
  {A.}~\bibnamefont {Kar}},\ }\href {https://doi.org/10.1063/1.2748328}
  {\bibfield  {journal} {\bibinfo  {journal} {Appl. Phys. Lett.}\ }\textbf
  {\bibinfo {volume} {90}},\ \bibinfo {pages} {241107} (\bibinfo {year}
  {2007})}\BibitemShut {NoStop}%
\bibitem [{\citenamefont {Rodenas}\ and\ \citenamefont
  {Kar}(2011)}]{multiscan20011oe}%
  \BibitemOpen
  \bibfield  {author} {\bibinfo {author} {\bibfnamefont {A.}~\bibnamefont
  {Rodenas}}\ and\ \bibinfo {author} {\bibfnamefont {A.~K.}\ \bibnamefont
  {Kar}},\ }\href {https://doi.org/10.1364/OE.19.017820} {\bibfield  {journal}
  {\bibinfo  {journal} {Opt. Express}\ }\textbf {\bibinfo {volume} {19}},\
  \bibinfo {pages} {17820} (\bibinfo {year} {2011})}\BibitemShut {NoStop}%
\bibitem [{\citenamefont {Staudt}\ \emph {et~al.}(2007)\citenamefont {Staudt},
  \citenamefont {Hastings-Simon}, \citenamefont {Nilsson}, \citenamefont
  {Afzelius}, \citenamefont {Scarani}, \citenamefont {Ricken}, \citenamefont
  {Suche}, \citenamefont {Sohler}, \citenamefont {Tittel},\ and\ \citenamefont
  {Gisin}}]{tittel2007prl}%
  \BibitemOpen
  \bibfield  {author} {\bibinfo {author} {\bibfnamefont {M.~U.}\ \bibnamefont
  {Staudt}}, \bibinfo {author} {\bibfnamefont {S.~R.}\ \bibnamefont
  {Hastings-Simon}}, \bibinfo {author} {\bibfnamefont {M.}~\bibnamefont
  {Nilsson}}, \bibinfo {author} {\bibfnamefont {M.}~\bibnamefont {Afzelius}},
  \bibinfo {author} {\bibfnamefont {V.}~\bibnamefont {Scarani}}, \bibinfo
  {author} {\bibfnamefont {R.}~\bibnamefont {Ricken}}, \bibinfo {author}
  {\bibfnamefont {H.}~\bibnamefont {Suche}}, \bibinfo {author} {\bibfnamefont
  {W.}~\bibnamefont {Sohler}}, \bibinfo {author} {\bibfnamefont
  {W.}~\bibnamefont {Tittel}},\ and\ \bibinfo {author} {\bibfnamefont
  {N.}~\bibnamefont {Gisin}},\ }\href
  {https://doi.org/10.1103/PhysRevLett.98.113601} {\bibfield  {journal}
  {\bibinfo  {journal} {Phys. Rev. Lett.}\ }\textbf {\bibinfo {volume} {98}},\
  \bibinfo {pages} {113601} (\bibinfo {year} {2007})}\BibitemShut {NoStop}%
\bibitem [{\citenamefont {Zhong}\ \emph {et~al.}(2016)\citenamefont {Zhong},
  \citenamefont {Rochman}, \citenamefont {Kindem}, \citenamefont {Miyazono},\
  and\ \citenamefont {Faraon}}]{zhong2016high}%
  \BibitemOpen
  \bibfield  {author} {\bibinfo {author} {\bibfnamefont {T.}~\bibnamefont
  {Zhong}}, \bibinfo {author} {\bibfnamefont {J.}~\bibnamefont {Rochman}},
  \bibinfo {author} {\bibfnamefont {J.~M.}\ \bibnamefont {Kindem}}, \bibinfo
  {author} {\bibfnamefont {E.}~\bibnamefont {Miyazono}},\ and\ \bibinfo
  {author} {\bibfnamefont {A.}~\bibnamefont {Faraon}},\ }\href
  {https://doi.org/10.1364/OE.24.000536} {\bibfield  {journal} {\bibinfo
  {journal} {Opt. Express}\ }\textbf {\bibinfo {volume} {24}},\ \bibinfo
  {pages} {536} (\bibinfo {year} {2016})}\BibitemShut {NoStop}%
\bibitem [{\citenamefont {Lauritzen}\ \emph {et~al.}(2012)\citenamefont
  {Lauritzen}, \citenamefont {Timoney}, \citenamefont {Gisin}, \citenamefont
  {Afzelius}, \citenamefont {de~Riedmatten}, \citenamefont {Sun}, \citenamefont
  {Macfarlane},\ and\ \citenamefont {Cone}}]{lauritzen2012spectroscopic}%
  \BibitemOpen
  \bibfield  {author} {\bibinfo {author} {\bibfnamefont {B.}~\bibnamefont
  {Lauritzen}}, \bibinfo {author} {\bibfnamefont {N.}~\bibnamefont {Timoney}},
  \bibinfo {author} {\bibfnamefont {N.}~\bibnamefont {Gisin}}, \bibinfo
  {author} {\bibfnamefont {M.}~\bibnamefont {Afzelius}}, \bibinfo {author}
  {\bibfnamefont {H.}~\bibnamefont {de~Riedmatten}}, \bibinfo {author}
  {\bibfnamefont {Y.}~\bibnamefont {Sun}}, \bibinfo {author} {\bibfnamefont
  {R.}~\bibnamefont {Macfarlane}},\ and\ \bibinfo {author} {\bibfnamefont
  {R.}~\bibnamefont {Cone}},\ }\href
  {https://doi.org/10.1103/PhysRevB.85.115111} {\bibfield  {journal} {\bibinfo
  {journal} {Phys. Rev. B}\ }\textbf {\bibinfo {volume} {85}},\ \bibinfo
  {pages} {115111} (\bibinfo {year} {2012})}\BibitemShut {NoStop}%
\bibitem [{\citenamefont {Macfarlane}\ \emph {et~al.}(2014)\citenamefont
  {Macfarlane}, \citenamefont {Arcangeli}, \citenamefont {Ferrier},\ and\
  \citenamefont {Goldner}}]{goldner2014prl}%
  \BibitemOpen
  \bibfield  {author} {\bibinfo {author} {\bibfnamefont {R.~M.}\ \bibnamefont
  {Macfarlane}}, \bibinfo {author} {\bibfnamefont {A.}~\bibnamefont
  {Arcangeli}}, \bibinfo {author} {\bibfnamefont {A.}~\bibnamefont {Ferrier}},\
  and\ \bibinfo {author} {\bibfnamefont {P.}~\bibnamefont {Goldner}},\ }\href
  {https://doi.org/10.1103/PhysRevLett.113.157603} {\bibfield  {journal}
  {\bibinfo  {journal} {Phys. Rev. Lett.}\ }\textbf {\bibinfo {volume} {113}},\
  \bibinfo {pages} {157603} (\bibinfo {year} {2014})}\BibitemShut {NoStop}%
\bibitem [{\citenamefont {Brendel}\ \emph {et~al.}(1999)\citenamefont
  {Brendel}, \citenamefont {Gisin}, \citenamefont {Tittel},\ and\ \citenamefont
  {Zbinden}}]{tb1999prl}%
  \BibitemOpen
  \bibfield  {author} {\bibinfo {author} {\bibfnamefont {J.}~\bibnamefont
  {Brendel}}, \bibinfo {author} {\bibfnamefont {N.}~\bibnamefont {Gisin}},
  \bibinfo {author} {\bibfnamefont {W.}~\bibnamefont {Tittel}},\ and\ \bibinfo
  {author} {\bibfnamefont {H.}~\bibnamefont {Zbinden}},\ }\href
  {https://doi.org/DOI 10.1103/PhysRevLett.82.2594} {\bibfield  {journal}
  {\bibinfo  {journal} {Phys. Rev. Lett.}\ }\textbf {\bibinfo {volume} {82}},\
  \bibinfo {pages} {2594} (\bibinfo {year} {1999})}\BibitemShut {NoStop}%
\bibitem [{\citenamefont {Marcikic}\ \emph {et~al.}(2002)\citenamefont
  {Marcikic}, \citenamefont {de~Riedmatten}, \citenamefont {Tittel},
  \citenamefont {Scarani}, \citenamefont {Zbinden},\ and\ \citenamefont
  {Gisin}}]{tb2002pra}%
  \BibitemOpen
  \bibfield  {author} {\bibinfo {author} {\bibfnamefont {I.}~\bibnamefont
  {Marcikic}}, \bibinfo {author} {\bibfnamefont {H.}~\bibnamefont
  {de~Riedmatten}}, \bibinfo {author} {\bibfnamefont {W.}~\bibnamefont
  {Tittel}}, \bibinfo {author} {\bibfnamefont {V.}~\bibnamefont {Scarani}},
  \bibinfo {author} {\bibfnamefont {H.}~\bibnamefont {Zbinden}},\ and\ \bibinfo
  {author} {\bibfnamefont {N.}~\bibnamefont {Gisin}},\ }\href
  {https://doi.org/ARTN 06230810.1103/PhysRevA.66.062308} {\bibfield  {journal}
  {\bibinfo  {journal} {Phys. Rev. A}\ }\textbf {\bibinfo {volume} {66}},\
  \bibinfo {pages} {062308} (\bibinfo {year} {2002})}\BibitemShut {NoStop}%
\bibitem [{\citenamefont {Zhou}\ \emph
  {et~al.}(2015{\natexlab{b}})\citenamefont {Zhou}, \citenamefont {Huelga},
  \citenamefont {Li},\ and\ \citenamefont {Guo}}]{zzq2015lg}%
  \BibitemOpen
  \bibfield  {author} {\bibinfo {author} {\bibfnamefont {Z.-Q.}\ \bibnamefont
  {Zhou}}, \bibinfo {author} {\bibfnamefont {S.~F.}\ \bibnamefont {Huelga}},
  \bibinfo {author} {\bibfnamefont {C.-F.}\ \bibnamefont {Li}},\ and\ \bibinfo
  {author} {\bibfnamefont {G.-C.}\ \bibnamefont {Guo}},\ }\href
  {https://doi.org/ARTN 11300210.1103/PhysRevLett.115.113002} {\bibfield
  {journal} {\bibinfo  {journal} {Phys. Rev. Lett.}\ }\textbf {\bibinfo
  {volume} {115}},\ \bibinfo {pages} {113002} (\bibinfo {year}
  {2015}{\natexlab{b}})}\BibitemShut {NoStop}%
\bibitem [{\citenamefont {Marcikic}\ \emph {et~al.}(2003)\citenamefont
  {Marcikic}, \citenamefont {Riedmatten}, \citenamefont {Tittel}, \citenamefont
  {Zbinden},\ and\ \citenamefont {Gisin}}]{gisin2003nature}%
  \BibitemOpen
  \bibfield  {author} {\bibinfo {author} {\bibfnamefont {I.}~\bibnamefont
  {Marcikic}}, \bibinfo {author} {\bibfnamefont {H.}~\bibnamefont
  {Riedmatten}}, \bibinfo {author} {\bibfnamefont {W.}~\bibnamefont {Tittel}},
  \bibinfo {author} {\bibfnamefont {H.}~\bibnamefont {Zbinden}},\ and\ \bibinfo
  {author} {\bibfnamefont {N.}~\bibnamefont {Gisin}},\ }\href
  {https://doi.org/10.1038/nature01376} {\bibfield  {journal} {\bibinfo
  {journal} {Nature}\ }\textbf {\bibinfo {volume} {421}},\ \bibinfo {pages}
  {509} (\bibinfo {year} {2003})}\BibitemShut {NoStop}%
\bibitem [{\citenamefont {Specht}\ \emph {et~al.}(2011)\citenamefont {Specht},
  \citenamefont {N{\"o}lleke}, \citenamefont {Reiserer}, \citenamefont
  {Uphoff}, \citenamefont {Figueroa}, \citenamefont {Ritter},\ and\
  \citenamefont {Rempe}}]{bound2011nature}%
  \BibitemOpen
  \bibfield  {author} {\bibinfo {author} {\bibfnamefont {H.}~\bibnamefont
  {Specht}}, \bibinfo {author} {\bibfnamefont {C.}~\bibnamefont {N{\"o}lleke}},
  \bibinfo {author} {\bibfnamefont {A.}~\bibnamefont {Reiserer}}, \bibinfo
  {author} {\bibfnamefont {M.}~\bibnamefont {Uphoff}}, \bibinfo {author}
  {\bibfnamefont {E.}~\bibnamefont {Figueroa}}, \bibinfo {author}
  {\bibfnamefont {S.}~\bibnamefont {Ritter}},\ and\ \bibinfo {author}
  {\bibfnamefont {G.}~\bibnamefont {Rempe}},\ }\href
  {https://doi.org/10.1038/nature09997} {\bibfield  {journal} {\bibinfo
  {journal} {Nature}\ }\textbf {\bibinfo {volume} {473}},\ \bibinfo {pages}
  {190} (\bibinfo {year} {2011})}\BibitemShut {NoStop}%
\bibitem [{\citenamefont {Equall}\ \emph {et~al.}(1994)\citenamefont {Equall},
  \citenamefont {Sun}, \citenamefont {Cone},\ and\ \citenamefont
  {Macfarlane}}]{EuT2o}%
  \BibitemOpen
  \bibfield  {author} {\bibinfo {author} {\bibfnamefont {R.~W.}\ \bibnamefont
  {Equall}}, \bibinfo {author} {\bibfnamefont {Y.}~\bibnamefont {Sun}},
  \bibinfo {author} {\bibfnamefont {R.~L.}\ \bibnamefont {Cone}},\ and\
  \bibinfo {author} {\bibfnamefont {R.~M.}\ \bibnamefont {Macfarlane}},\ }\href
  {https://doi.org/10.1103/PhysRevLett.72.2179} {\bibfield  {journal} {\bibinfo
   {journal} {Phys. Rev. Lett.}\ }\textbf {\bibinfo {volume} {72}},\ \bibinfo
  {pages} {2179} (\bibinfo {year} {1994})}\BibitemShut {NoStop}%
\bibitem [{\citenamefont {Zhong}\ \emph
  {et~al.}(2015{\natexlab{b}})\citenamefont {Zhong}, \citenamefont {Hedges},
  \citenamefont {Ahlefeldt}, \citenamefont {Bartholomew}, \citenamefont
  {Beavan}, \citenamefont {Wittig}, \citenamefont {Longdell},\ and\
  \citenamefont {Sellars}}]{nature_6hour}%
  \BibitemOpen
  \bibfield  {author} {\bibinfo {author} {\bibfnamefont {M.}~\bibnamefont
  {Zhong}}, \bibinfo {author} {\bibfnamefont {M.~P.}\ \bibnamefont {Hedges}},
  \bibinfo {author} {\bibfnamefont {R.~L.}\ \bibnamefont {Ahlefeldt}}, \bibinfo
  {author} {\bibfnamefont {J.~G.}\ \bibnamefont {Bartholomew}}, \bibinfo
  {author} {\bibfnamefont {S.~E.}\ \bibnamefont {Beavan}}, \bibinfo {author}
  {\bibfnamefont {S.~M.}\ \bibnamefont {Wittig}}, \bibinfo {author}
  {\bibfnamefont {J.~J.}\ \bibnamefont {Longdell}},\ and\ \bibinfo {author}
  {\bibfnamefont {M.~J.}\ \bibnamefont {Sellars}},\ }\href
  {https://doi.org/10.1038/nature14025} {\bibfield  {journal} {\bibinfo
  {journal} {Nature}\ }\textbf {\bibinfo {volume} {517}},\ \bibinfo {pages}
  {177} (\bibinfo {year} {2015}{\natexlab{b}})}\BibitemShut {NoStop}%
\end{thebibliography}%

\end{document}